\documentclass{article}

\usepackage[a4paper]{geometry}
\usepackage{amsmath}
\usepackage{natbib}
\usepackage{amsfonts}
\usepackage{float}
\usepackage{amsthm}
\usepackage{graphicx,multirow}
\usepackage{amssymb}
\usepackage{color}
\usepackage[colorlinks]{hyperref}
\usepackage{hyperref}
\usepackage{xr}
\hypersetup{
	colorlinks,
	citecolor=black,
	filecolor=black,
	linkcolor=blue,
	urlcolor=blue
}

\usepackage{appendix}

\usepackage{titling}
\usepackage{pdfpages}
\title{Generalized Sparse Additive Models}
\author{Asad Haris\thanks{asad.haris@mail.mcgill.ca, Department of Epidemiology, Biostatistics and Occupational Health, McGill University} \and Noah Simon\thanks{nrsimon@uw.edu, Department of Biostatistics, University of Washington} \and Ali Shojaie\thanks{ashojaie@uw.edu, Department of Biostatistics, University of Washington}}

\usepackage{algorithm}
\usepackage[]{algpseudocode}

\newtheorem{theorem}{Theorem}
\newtheorem{corollary}{Corollary}
\newtheorem{lemma}{Lemma}

\newtheorem{definition}{Definition}

\newcommand{\bs}[1]{\boldsymbol{#1}}

\newcommand{\wt}[1]{\widetilde{#1}}
\newcommand{\wh}[1]{\widehat{#1}}

\newcommand{\e}{\varepsilon}

\newcommand{\name}{\texttt{GSAM}}

\begin{document}

\maketitle
\begin{abstract}
We present a unified framework for estimation and analysis of generalized additive models in high dimensions. The framework defines a large class of penalized regression estimators, encompassing many existing methods.
An efficient computational algorithm for this class is presented that easily scales to thousands of observations and features. We prove minimax optimal convergence bounds for this class under a weak compatibility condition. In addition, we characterize the rate of convergence when this compatibility condition is not met. Finally, we also show that the optimal penalty parameters for structure and sparsity penalties in our framework are linked, allowing cross-validation to be conducted over only a single tuning parameter. We complement our theoretical results with empirical studies comparing some existing methods within this framework.
\end{abstract}

\section{Introduction}
\label{sec:introduction}

In this paper, we model a response variable as an additive function of a potentially large number of covariates. The problem can be formulated as follows: we are given $n$ observations with response $y_i\in\mathbb{R}$ and covariates $\bs{x}_i\in \mathbb{R}^p$ for $i = 1, \ldots, n$. The goal is to fit the model
\[
g\left(\mathbb{E}\left(y_i|\bs{x}_i\right)\right) = \beta_0 + \sum_{j=1}^p f_j\left(x_{ij}\right), \quad i=1,\ldots, n,
\]
for a prespecified \emph{link} function $g$, unknown intercept $\beta_0$ and, unknown component functions $f_1,\ldots,f_p$. The link function, $g$, is generally based on the outcome data-type, e.g., $g(x) = x$ or $g(x) = \log(x)$ for continuous or count response data, respectively. The estimands, $f_1,\ldots,f_p$, give the conditional relationships between each feature $x_{ij}$ and the outcome $y_i$ for all $i$ and $j$. 
For identifiability, we assume $\sum_{i=1}^n f_j(x_{ij}) = 0$ for all $j=1,\ldots,p$. This model is known as a generalized additive model (GAM) \citep{hastie1990generalized}. It extends the generalized linear model~(GLM) where each $f_j$ is linear, 
and is a popular choice for modeling different types of response variables 
as a function of covariates. GAMs are popular because they extend GLMs to model non-linear conditional relationships while retaining some interpretability (we can examine the effect of each covariate  $x_{ij}$ individually on $y_i$ while holding all other variables fixed); they also do not suffer from the \emph{curse of dimensionality}. 


While there are a number of proposals for estimating GAMs, a popular approach is to encode the estimation in the following convex optimization problem~\citep{sadhanala2017additive}:
\begin{equation}\label{eq:add}
\wh{\beta}, \wh{f}_1,\ldots,\wh{f}_p \gets \underset{\beta\in \mathbb{R}, f_1,\ldots,f_p\in \mathcal{F}}{\operatorname{argmin}}\  -n^{-1}\sum_{i=1}^n \ell\Big(y_i, \beta + \sum_{j=1}^p f_j\left(x_{ij}\right)\Big) + \lambda_{st}\sum_{j=1}^p P_{st}\left(f_j\right).
\end{equation}
Here $\mathcal{F}$ is some suitable function class; $\ell(y_i,\theta)$ is the log-likelihood of $y_i$ under parameter $\theta$; $P_{st}$ is a structure-inducing penalty to control the wildness of the estimated functions, $\wh{f}_j$; and $\lambda_{st} > 0$ is a penalty parameter which modulates the trade-off between goodness-of-fit and structure/smoothness of estimates. 
The class $\mathcal{F}$ is a general convex space, e.g., $\mathcal{F} = L^2[0,1]$. Functions $-\ell(y_i,\theta)$ and $P_{st}(f_j)$ are convex in $\theta$ and $f_j$, respectively. The objective function in~\eqref{eq:add} is convex and for small dimension, $p$, can be solved via a general-purpose convex solver. However, many modern datasets are high-dimensional, often with more features than observations, i.e., $p > n$. Fitting even GLMs is challenging in such settings as conventional methods are known to overfit the data. 
A common assumption in the high-dimensional setting is \emph{sparsity}, that is, only a small (but unknown) subset of features is informative for the outcome. In this case, it is desirable to apply feature selection: to build a model for which only a small subset of $\wh{f}_j \not\equiv 0$.

A number of estimators have been proposed for fitting GAMs with sparsity. These estimators are generally solutions to a convex optimization problem.
Though they differ in details, we show that most of these optimization problems can be written as:
\begin{equation}\label{eq:sparse-add}
\small
\wh{\beta}, \wh{f}_1,\ldots,\wh{f}_p  \gets \hspace{-4mm} \underset{\beta\in \mathbb{R}, f_1,\ldots,f_p \in \mathcal{F}}{\operatorname{argmin}} \hspace{-2mm}  -n^{-1}\sum_{i=1}^n \ell\Big(y_i, \beta + \sum_{j=1}^p f_j\left(x_{ij}\right)\Big) + \lambda_{st}\sum_{j=1}^p P_{st}\left(f_j\right) + \lambda_{sp}\sum_{j=1}^p\left\|f_j\right\|_n,
\end{equation}
where $\|f_j\|_n = \Big[n^{-1}\sum_{i=1}^{n}\{f_j(x_{ij})\}^2\Big]^{1/2}$ is a group lasso-type penalty \citep{yuan2006model} for feature-wise sparsity, and $\lambda_{sp}$ a sparsity-related tuning parameter \citep{ravikumar2009sparse, lou2016sparse, petersen2016fused, sadhanala2017additive, koltchinskii2010sparsity, raskutti2012minimax, yuan2015minimax, meier2009high}. However, previous proposals consists of gaps around efficient computation~\citep{koltchinskii2010sparsity, raskutti2012minimax, yuan2015minimax} and/or optimal statistical convergence properties~\citep{ravikumar2009sparse, lou2016sparse, petersen2016fused, sadhanala2017additive}. General-purpose convex solvers have also been suggested~\citep{koltchinskii2010sparsity, raskutti2012minimax, yuan2015minimax}  as an alternative for solving problem \eqref{eq:sparse-add}, but they roughly scale as $O(n^3p^3)$ and are hence inefficient. This manuscript aims to bridge these gaps.

We present a general framework for sparse GAMs with two major contributions, a general algorithm for computing~\eqref{eq:sparse-add} and a theorem for establishing convergence rates. Briefly, our algorithm is based on accelerated proximal gradient descent. This reduces \eqref{eq:sparse-add} to repeatedly solving a univariate penalized least squares problem. In many cases, this algorithm has a per-iteration complexity of $O(np)$ --- precisely that of state-of-the-art algorithms for the lasso \citep{friedman2010regularization,beck2009gradient}.  Our main theorem establishes fast convergence rates of the form $\max( s\log p/n, s\xi_n )$, where $s$ is the number of signal variables and $\xi_n$ is the minimax rate of the univariate regression problem, i.e., problem \eqref{eq:add} with $p=1$. Nonparametric rates are established for a wide class of structural penalties $P_{st}$ with $\xi_n = n^{-2m/(2m+1)}$, popular choices of $P_{st}$ include $m$-th order Sobolev and H\"{o}lder norms, total variation norm of the $m$-th derivative and, norms of Reproducing Kernel Hilbert Spaces (RKHS). Parametric rates are also established with $\xi_n = T_n/n$ via a truncation-penalty; the number of parameters, $T_n$, can be fixed or allowed to grow with sample size. 

The highlight of this paper is the generality of the proposed framework: not only does it encompass many existing estimators for high-dimensional GAMs, but also estimators for low-dimensional GAMs, low-dimensional fully nonparametric models and, parametric models in low or high-dimensional settings. 
As a byproduct of our general theorem, we also determine that $\lambda_{st} = \lambda_{sp}^2$ in \eqref{eq:sparse-add} results in optimal convergence rates, reducing the problem to a single tuning parameter. 

The rest of the paper is organized as follows. In Section~\ref{sec:Methodology}, we detail our framework and discuss various choices of structural penalties, $P_{st}$, illustrating that our framework encompasses many existing proposals. In Section~\ref{sec:Algorithm} we present an algorithm for solving the optimization problem \eqref{eq:sparse-add} for a broad class of $P_{st}$ penalties, and establish their theoretical convergence rates in Section~\ref{sec:TheoreticalResults}. We explore the empirical performance of various choices of $P_{st}$ in simulation in Section~\ref{sec:SimulationStudy}, and in an application to the Boston housing dataset in Section~\ref{sec:data}. Concluding remarks are in Section~\ref{sec:Conclusion}.

\section{General Framework for Additive Models}
\label{sec:Methodology}

In this section, we present our general framework for estimating sparse GAMs, discuss its salient features, and review some existing methods as special cases.  
Before presenting our framework, we introduce some notation. For any function $f$ and response/covariate pair, $(y,\bs{x})$, let $-\ell(f) \equiv -\ell(y,f({\bs{x}}))$ denote a loss function; given data $(y_1,\bs{x}_1), \ldots, (y_n, \bs{x}_n)$, let $\mathbb{P}_n \ell(f) \equiv n^{-1} \sum_{i=1}^n \ell(y_i,f({\bs{x}}_i))$ denote an empirical average; and $\|f\|_n^2 \equiv n^{-1}\sum_{i=1}^n f({\bs{x}}_i)^2$ denote the empirical norm. With some abuse of notation, we will use the shorthand $f_j$ to denote the function $f_j\circ \pi_j$ where $\pi_j(\bs{x})= x_j$ for $\bs{x}\in\mathbb{R}^p$. 

Our general framework for obtaining a \emph{\underline{G}eneralized \underline{S}parse \underline{A}dditive \underline{M}odel} (\name) encompasses estimators that can be obtained by solving the following problem:
\begin{equation} \label{eq:unif-sparse-add1}
\wh{\beta}, \wh{f}_1,\ldots,\wh{f}_p \gets \hspace{-3mm} \underset{\beta\in \mathbb{R},f_1,\ldots,f_p \in \mathcal{F}}{\operatorname{argmin}}  \underbrace{-\mathbb{P}_n \ell\left(\beta + \sum_{j=1}^p f_j\right)}_{\text{Goodness-of-fit}} + \underbrace{\lambda^2\sum_{j=1}^p P_{st}\left(f_j\right)}_{\text{structure-inducing}} +  \underbrace{\lambda\sum_{j=1}^p\left\|f_j\right\|_n}_{\text{sparsity-inducing}}.
\end{equation} 
This optimization problem balances three terms. The first is a loss function based on goodness-of-fit to the observed data; the least squares loss, $-\ell(f) = (y - f(\bs{x}))^2$, is commonly used for continuous response. Our general framework requires only convexity and differentiability of $-\ell(y, \theta)$, with respect to $\theta$. Later we consider loss functions given by the negative log-likelihood of exponential family distributions. The second piece is a penalty to induce smoothness/structure of the function estimates. Our framework requires $P_{st}$ to be a \emph{semi-norm} on $\mathcal{F}$. This choice is motivated by both statistical theory and computational efficiency; we discuss this along with possible choices of $P_{st}$ in the following sub-sections. The final piece is a sparsity penalty $\|\cdot\|_n$, which encourages models with $\wh{f}_j \equiv 0$ for many $j$. Surprisingly, $P_{st}$ also plays an important role in obtaining an appropriate sparsity pattern. Briefly, if $P_{st}$ is a squared semi-norm then either all $\wh{f}_j\equiv0$ or all $\wh{f}_j\not\equiv 0$. To fit models where some $\wh{f}_j\equiv0$ and not others, the non-differentiablity of semi-norms at 0 is crucial, we detail this in Section~\ref{sec:Insights} below. Throughout this manuscript, we require the function class $\mathcal{F}$ to be a convex cone, e.g., $L^2(\mathbb{R})$. Later for some specific results, we will additionally require $\mathcal{F}$ to be a linear space.

As noted before, the tuning parameters for structure ($\lambda$) and sparsity ($\lambda^2$) are coupled in our framework. The theoretical consequence of this is that, for properly chosen $\lambda$, we get rate-optimal estimates (shown in Section~\ref{sec:TheoreticalResults}). The practical consequence is that we have a single tuning parameter. This is adequate for most choices of $P_{st}$ as seen in our empirical experiments of Section~\ref{sec:SimulationStudy}.

Furthermore, our framework relaxes the usual distributional requirements of i.i.d. response from an exponential family; we require only $y_i$ independent and $E\{y_i - E(y_i)\}$ to be sub-Gaussian~(or sub-Exponential). This demonstrates the generality of our framework and highlights our main innovation: the efficient algorithm of Section~\ref{sec:Algorithm} and theoretical results of Section~\ref{sec:TheoreticalResults} apply to a very broad class of estimators, fill in the gaps of existing work and, can easily be applied for the development of future estimators.

\subsection{Structure Inducing Penalties}
We now present some possible choices of the structural penalty $P_{st}$ followed by a discussion of the conditions on $P_{st}$ that lead to desirable estimation and computation. The main requirement is that $P_{st}$ is a semi-norm: a functional that obeys all the rules of a norm except one --- for nonzero $f$ we may have $P_{st}(f) = 0$. 
Some potential choices for smoothing semi-norms are:
\begin{enumerate}
\item $k$-th order Sobolev  \hfill $P_{st}\gets P_{sobolev}(f^{(k)}) = \sqrt{\int_x \left\{f^{(k)}(x)\right\}^2 dx}$;
\item $k$-th order total variation \hfill $P_{st}\gets {TV}(f^{(k)})$;
\item $k$-th order H\"{o}lder  \hfill $P_{st}\gets P_{holder}(f^{(k)}) = \operatorname{sup}_x \left| f^{(k)}(x) \right|$;
\item 
$k$-th order monotonicity \hfill $P_{st}\gets P_{mon}(f^{(k)}) \gets \mathbb{I}\left(f;\  \{f:f^{(k+1)}\ge 0\}\right)$;
\item 
$M$-th dimensional linear subspace \hfill $P_{st}\gets P^M_{lin}(f) = \mathbb{I}\left(f;\ \textrm{span}\left\{g_1,\ldots, g_M\right\}\right)$;
\end{enumerate}
here $TV(\cdot)$ is the total variation norm, $TV(f) = \sup \{\sum_{i=1}^{o}|f(z_{i+1}) - f(z_i)|: z_1<\ldots<z_o \text{ is a partition of }[0,1] \}$, and $\mathbb{I}$ is a convex indicator function defined as $\mathbb{I}(f;\mathcal{A}) = 0$ if $f\in \mathcal{A}$ and $\mathbb{I}(f;\mathcal{A}) = \infty$ if $f\not\in \mathcal{A}$. As implied by the name, $P_{st}$ imposes smoothness or structure on individual components $\wh{f}_j$. For instance, $P_{sobolev}(f^{\prime\prime})$ is a common measure of smoothness; small $\lambda$ values leads to wiggly fitted functions $\wh{f}_j$; on the other hand, sufficiently large $\lambda$ values would lead to each component being a  linear function. The convex indicator function, $\mathbb{I}(\cdot)$, can impose specific structural properties on $\wh{f}_j$; e.g., $P_{mon}(f)$ fits a model with each $\wh{f}_j$ a non-decreasing function.

The semi-norm requirement for $P_{st}$ is important because: (a) it implies convexity leading to a convex objective function, (b) the first order  absolute homogeneity ($P_{st}(\alpha f) = |\alpha|P_{st}(f)$) is needed for the algorithm of Section~\ref{sec:Algorithm} and, (c) the triangle inequality is used throughout the proof of our theoretical results of Section~\ref{sec:TheoreticalResults}. For our context, we consider convex indicators of cones as a semi-norm because, the first order homogeneity condition can be relaxed. For our algorithm, we only require $P_{st}(\alpha f) = \alpha P_{st}(f)$ for $\alpha>0$; for our theoretical results we treat convex indicators of cones as a special case and discuss them at the end of Section~\ref{sec:main}. For non-sparse GAMs of the form \eqref{eq:add}, the existing literature does not necessarily use a semi-norm penalty; a common choice of smoothing penalty is $P_{st}(f) = P^2_{sobolev}(f^{\prime\prime})$. In the following subsection, we discuss the issues with using squared semi-norm penalties in high dimensions, particularly their impact on the sparsity of estimated component functions.  

\subsection{Semi-norms vs Squared Semi-norms}
\label{sec:Insights}
Given a semi-norm $P_{semi}$, using $P_{st} = P_{semi}^2$ in \eqref{eq:unif-sparse-add1} may give poor theoretical performance (as noted in \cite{meier2009high} for $P_{semi} = P_{sobolev}$) and, can also be computationally expensive (as disscussed in Section~\ref{sec:Algorithm}). In this subsection, we show a surprising result: using a squared semi-norm penalty does not actually lead to a sparse solution. 

To be precise, using $P_{st} = P_{semi}^2$ leads to an active set ${S} = \{j: \wh{f}_j \not\equiv 0\}$, for which either $|{S}| = 0$ or $|{S}| = p$; in contrast, using $P_{st} = P_{semi}$ can give active sets such that $0<|{S}|<p$. To demonstrate this phenomenon, we consider first the univariate problem 
\begin{align}
\label{eq:univ1}
\wh{f}_1 \gets \underset{f\in \mathcal{F}}{\operatorname{argmin}}\  \frac{1}{n}\sum_{i=1}^n \left(y_i - f\left(x_{i}\right)\right)^2 + \lambda_{st} P^2_{semi}\left(f\right) + \lambda_{sp}\left\|f\right\|_n,
\end{align}
and characterize conditions for which $\wh{f}_1\equiv 0$. Recall that $\wh{f}_1$ minimizes the objective in \eqref{eq:univ1} if for every direction $h$, the objective is minimized at $\e = 0$ along the path $\wh{f}_1 + \e h$. The following lemma gives necessary and sufficient conditions for $\wh{f}_1$ to be $\bs{0}$. 
\begin{lemma}
	\label{lemma:univ1}
	For $\wh{f}_1$ given by \eqref{eq:univ1}, the following are equivalent: (a) $\wh{f}_1 = \bs{0}$, (b) for every direction $h\in \mathcal{F}$, $\left|{n^{-1}}\sum_i y_i {h(x_i)}/{\left\|h\right\|_n}\right| \leq \lambda_{sp}$, (c) $\|\bs{y}\|_n\le \lambda_{sp}$.
\end{lemma}
Lemma~\ref{lemma:univ1} is proved in Appendix~\ref{sec:ProofSec2} in the supplementary material. 
Condition (c) of Lemma~\ref{lemma:univ1} is problematic when we consider multiple features in our additive problem \eqref{eq:unif-sparse-add1}. For additive models, condition (c) implies that sparsity of component $\wh{f}_j$, does not depend on covariate $j$. Thus if all smoothing penalties are squared semi-norms then for a given $\lambda_{sp}$, there exists a minimizer with  either all $\wh{f}_j = {\bf 0}$ or all $\wh{f}_j\not = {\bf 0}$. 
Consider, instead, the optimization problem 
\begin{equation}\label{eq:univ2}
\wh{f}_2 \gets \underset{f\in \mathcal{F}}{\operatorname{argmin}}\  \frac{1}{n}\sum_{i=1}^n \left(y_i - f\left(x_{i}\right)\right)^2 + \lambda_{st} P_{semi}\left(f\right) + \lambda_{sp}\left\|f\right\|_n.
\end{equation}
For this problem, we obtain the following result~(proof in Appendix~\ref{sec:ProofSec2} in the supplementary material).

\begin{lemma}
	\label{lemma:univ2}
	For $\wh{f}_2$ defined by \eqref{eq:univ2}, the following are equivalent: (a) $\wh{f}_2 = \bs{0}$, (b) for every direction $h$, there exists some $V\in [-1,1]$ such that $\Big|{n^{-1}}\sum_i y_i {h(x_i)}/{\|h\|_n} - \lambda_{st}{VP_{semi}(h)}/{\|h \|_n}\Big| \leq \lambda_{sp}$. 
	Additionally, if $\|\bs{y}\|_n \le \lambda$ then $\wh{f}_2=\bs{0}$, but the converse is not necessarily true. 
\end{lemma}
Unlike the squared semi-norm penalties, conditions for $\wh{f}_2 = \bs{0}$ involve the feature vector $\bs{x}$. Thus for an additive model the sparsity of component $j$ depends on both the response vector $\bs{y}$, and $j$-th covariate $(x_{1j}, \ldots, x_{nj})$. Consequently, there are many $(\lambda_{sp}, \lambda_{st})$ pairs for which we will have some $\wh{f}_j = {\bf 0}$ and some $\wh{f}_j \not = {\bf 0}$. Additionally, Lemma~\ref{lemma:univ2} gives us a conservative value for $\lambda_{max} = \|\bs{y}\|_n$, i.e., the $\lambda_{sp}$ value for which all $\wh{f}_j = \bs{0}$.

\subsection{Relationship of Existing Methods to \name}
\label{sec:prevWork}
We now discuss some of the existing methods for sparse additive models in greater detail, and demonstrate that many existing proposals are special cases of our \name\ framework. One of the first proposals for sparse additive models, SpAM \citep{ravikumar2009sparse}, uses a basis expansion and solves
\begin{equation}\label{eq:SpAM}
\underset{\beta_1,\ldots,\beta_j\in \mathbb{R}^M}{\operatorname{argmin}}\   \Big\|{\bs{y}} - \sum_{j=1}^p \sum_{m=1}^M \beta_{jm}\bs{\psi}_{jm}\Big\|_n^2 + \lambda\sum_{j=1}^p \Big\|\sum_{m=1}^M \beta_{jm}\bs{\psi}_{jm}\Big\|_n,
\end{equation}
where $\bs{\psi}_{jm} = [\psi_m(x_{1j}),\ldots, \psi_m(x_{nj})]^T\in \mathbb{R}^n$ for basis functions $\psi_1,\ldots,\psi_M$. This is a  \name\ with $P_{st} = \mathbb{I}\left(f;\textrm{span}\left\{\psi_1,\ldots, \psi_M\right\}\right)$. The SpAM proposal is extended to partially linear models in SPLAM \citep{lou2016sparse}. There, a similar basis expansion is used, though with the particular choice $\psi_1(x) = x$. The SPLAM estimator solves
\begin{equation}\label{eq:SPLAM}
\small
\underset{\beta_1,\ldots,\beta_j\in \mathbb{R}^M}{\operatorname{argmin}}\  \Big\|{\bs{y}} - \sum_{j=1}^p \sum_{m=1}^M \beta_{jm}\bs{\psi}_{jm}\Big\|_n^2 + \lambda_1\sum_{j=1}^p \Big\|\sum_{m=1}^M \beta_{jm}\bs{\psi}_{jm}\Big\|_n + \lambda_2\sum_{j=1}^p \Big\|\sum_{m=2}^M \beta_{jm}\bs{\psi}_m\Big\|_n,
\end{equation}
and is also a \name\ with 
$$P_{st} = \mathbb{I}\left(f;\textrm{span}\left\{\psi_1,\ldots, \psi_M\right\}\right) + \sum_{j=1}^p \Big\|\operatorname{Proj}_{\operatorname{span}\left(\psi_2,\ldots,\psi_M\right)}\left(f\right)\Big\|_n,$$
where $\operatorname{Proj}_{A}$ is the projection operator onto the set $A$. The recently proposed extensions of trend filtering to additive models are other examples \citep{petersen2016fused,sadhanala2017additive}; these methods can be written in our \name\ framework with $P_{st}(f) = TV(f)$. 

\cite{koltchinskii2010sparsity}, \citet{raskutti2012minimax} and \cite{yuan2015minimax} discuss a similar framework to \name s; however, they only consider structural penalties $P_{st}$, which are norms of Reproducing Kernel Hilbert Spaces~(RKHS). Furthermore, they do not discuss efficient algorithms for solving the convex optimization problem. Using properties of RKHS, they note that their estimator is the minimum of a $d = np$ dimensional second order cone program (SOCP). The computation for general-purpose SOCP solvers scales roughly as $d^3$. Thus for even moderate $p$ and $n$, these problems quickly become intractable.

\citet{meier2009high} give two proposals: the first solves the optimization problem
\begin{equation*}
\underset{f_1,\ldots,f_p \in \mathcal{F}}{\operatorname{argmin }} \    \Big\|\bs{y} - \sum_{j=1}^{p}f_j \Big\|_n^2 + \sum_{j=1}^p \lambda_{sp}\sqrt{\left\|f_j\right\|^2_n + \lambda_{st} P_{st}^2\left(f_j\right)},
\end{equation*}
and is not a \name; they note that this proposal gives a suboptimal rate of convergence. The second is a \name\ of the form \eqref{eq:unif-sparse-add1} with $P_{st}(f) = P_{sobolev}(f^{\prime\prime})$. At the time, \cite{meier2009high} focused on the first proposal as no computationally efficient method for solving the second one was known to them. In a follow-up paper, \cite{geer2010lasso} studied the theoretical properties of a \name\ with an alternative, \emph{diagonalized smoothness} structural penalty. The diagnolized smoothness penalty for a function with basis expansion $f_{\bs{\beta}}(x) = \sum_{j=1}^{n}\psi_j(x)\beta_j$, is defined as 
\begin{equation}
\label{eqn:diagSmoothness}
P_{st}(f_{\bs{\beta}}) =  \Big( {\sum_{j=1}^{n}j^{2m}\beta_j^2} \Big)^{1/2},
\end{equation}
for a smoothness parameter $m$. All of the above mentioned proposals either fail to provide an efficient computational algorithm or have sub-optimal convergence rates.   
There are also a number of other proposals that do not quite fall in the \name\ framework \citep{chouldechova2015generalized, fan2012nonparametric, yin2012group}.


\section{General-Purpose Algorithm}
\label{sec:Algorithm}

Here we give a general algorithm for fitting \name s based on proximal gradient descent~\citep{parikh2014proximal}. We begin with some notation. We denote by $\dot{\ell}(y,\theta)$ and $\ddot{\ell}(y,\theta)$ the first and second derivatives of $\ell$ with respect to $\theta$. For functions $f, g:\mathbb{R}^p\to \mathbb{R}$, let $\langle f, \dot{\ell}(g)\rangle_n \equiv {n^{-1}}\sum_{i=1}^n f(\bs{x}_i) \{ \dot{\ell}(y_i,g(\bs{x}_i))\}$, $\mathbb{P}_n{\dot{\ell}}(g) \equiv {n^{-1}}\sum_{i=1}^n \dot{\ell}(y_i,g(\bs{x}_i) )$ and, $\|f + \dot{\ell}(g)\|_n^2 \equiv {n^{-1}}\sum_{i=1}^n \{ f(\bs{x}_i) + \dot{\ell}(y_i,g(\bs{x}_i) )\}^2$.
 
We begin with a second order Taylor expansion of the loss.
For this, we first apply Taylor's theorem to $\ell(y_i, \beta + \theta_{i1}+\ldots+\theta_{ip})$ as a $(p+1)$ variate function of $(\beta, \theta_{i1},\ldots,\theta_{ip})$. Note that for $|\ddot{\ell}(y,\theta)| \leq L$, the Hessian matrix, $H_{p+1}$, of $\ell(y_i, \beta + \theta_{i1}+\ldots+\theta_{ip})$ obeys the inequality $\bs{a}^TH_{p+1}\bs{a} \le (p+1)L\|a\|_2^2$ for all $\bs{a}\in \mathbb{R}^{p+1}$~\citep{zhan2005extremal}. This gives us the following bound:
 \begin{align*}
-\mathbb{P}_n\ell\Big(\beta+\sum_{j=1}^p f_j\Big) &\leq  -\mathbb{P}_n\ell\Big(\beta^0+\sum_{j=1}^p f^0_j\Big)  \\
&- (\beta - \beta^0)\mathbb{P}_n\dot{\ell}\Big(\beta^0 + \sum_{j=1}^p f^0_j\Big) - \sum_{j=1}^p \Big\langle f_j - f^0_j,\dot{\ell}\Big(\beta^0 + \sum_{j=1}^p f^0_j\Big)\Big\rangle_n   \\
&+ \frac{(p+1)L}{2}(\beta - \beta^0)^2 + \sum_{j=1}^p \frac{(p+1)L}{2}\left\|f_j - f^0_j\right\|_n^2, 
\end{align*}
which leads to the  following majorizing inequality
\begin{equation}
\begin{split}
-\mathbb{P}_n\ell\Big(\beta+\sum_{j=1}^p f_j\Big) &\leq
\frac{(p+1)L}{2}\Big[ \beta - \Big\{  \beta^0 + \frac{1}{(p+1)L}\mathbb{P}_n\dot{\ell}\Big(\beta^0 + \sum_{j=1}^p f^0_j\Big)  \Big\} \Big]^2 \\
&+ \sum_{j=1}^p \frac{(p+1)L}{2}\Big\|f_j - \Big\{ f^0_j +\frac{1}{(p+1)L} \dot{\ell}\Big(\beta^0+\sum_{j=1}^p f^0_j\Big) \Big\}\Big\|_n^2 + W  ,
\end{split}
\label{eq:major}
\end{equation}
where $W$ is not a function of $\beta$ or $f_j$ for any $j$.
Instead of minimizing the original problem~\eqref{eq:unif-sparse-add1}, we minimize the majorizing surrogate
\begin{align}\label{eq:prox}
\nonumber
\frac{1}{2} \Big[ \beta - \Big\{  \beta^0 + {t}\mathbb{P}_n\dot{\ell}\Big(\beta^0 + \sum_{j=1}^p f^0_j\Big)  \Big\} \Big]^2 &+ \frac{1}{2}\sum_{j=1}^p \Big\|f_j - \Big\{ f^0_j +{t} \dot{\ell}\Big(\beta^0 +\sum_{j=1}^p f^0_j\Big) \Big\}\Big\|_n^2 \\ 
&+  {t\lambda^2} \sum_{j=1}^p P_{st}\left(f_j\right) + {t\lambda} \sum_{j=1}^p \left\|f_j\right\|_n,
\end{align}
where $t = \{(p+1)L\}^{-1}$.
Minimizing \eqref{eq:prox} and re-centering our Taylor series at the current iterate, is precisely the proximal gradient recipe. 
Updating the intercept $\beta$, is simply $\wh{\beta}\gets \beta^0 + t\mathbb{P}_n\dot{\ell}\Big(\beta^0 + \sum_{j=1}^p f^0_j\Big)$. Components $f_1,\ldots, f_p$, can be updated in parallel by solving the univariate problems:
 \begin{equation}\label{eq:univ-emp}
\wh{f}_j \gets \underset{f\in \mathcal{F}}{\operatorname{argmin}}\ \frac{1}{2} \Big\|\Big\{ f^0_j + t\dot{\ell}\Big(\beta^0 + \sum_{j=1}^p f^0_j\Big) \Big\} - f\Big\|_n^2 + t{\lambda^2}P_{st}\left(f\right) + t{\lambda} \left\|f\right\|_n .
 \end{equation}
 At first, this problem still appears difficult due to the combination of structure and sparsity penalties. However, the following Lemma shows that things greatly simplify.
 
\begin{lemma}
\label{lemma:sofThres}
 Suppose $P_{st}$ is a semi-norm, and $\bs{r}$ is an $n$-vector. Consider the optimization problems 
  \begin{align}
 &\underset{f\in \mathcal{F}}{\operatorname{argmin}}\  \frac{1}{2} \|\bs{r} - f\|_n^2 + \lambda_1 P_{st}\left(f\right) + \lambda_2 \|f\|_n \label{eq:opt1*} ,\\
 &\underset{f\in \mathcal{F}}{\operatorname{argmin}}\  \frac{1}{2} \| \bs{r} - f\|_n^2 + \lambda_1 P_{st}\left(f\right) . \label{eq:opt2*}
  \end{align}
  If $\wt{f}$ is a solution to \eqref{eq:opt2*}; then $\wh{f}$ is a solution to \eqref{eq:opt1*} where $\wh{f}$ is defined as
  \begin{equation}\label{eq:softscale*}
  \wh{f} = \Big(1 - {\lambda_2}/{\|\wt{f}\|_n}\Big)_{+}\wt{f} ,
  \end{equation}
  with $(z)_+ = \max(z, 0)$.
\end{lemma}

The proof is given in Appendix~\ref{sec:appendix} in the supplementary material. Using Lemma~\ref{lemma:sofThres}, we can get the solution to \eqref{eq:univ-emp} by solving a problem in the form of \eqref{eq:opt2*}, a classical univariate smoothing problem, and then applying \eqref{eq:softscale*}, the simple soft-scaling operator. 
Putting things together, our proximal gradient algorithm for solving \eqref{eq:unif-sparse-add1} is summarized in Algorithm~\ref{alg:additive}.
 \begin{algorithm}
\caption{General Proximal Gradient Algorithm for \eqref{eq:unif-sparse-add1}}
\label{alg:additive}
\begin{algorithmic}[1]
\State Initialize $f^{0}_1,\ldots f^{0}_p \gets {\bf 0}, \beta^0 \gets 0$, $k\gets 1$; choose a step-size $t$
\While{$k\le max\_iter$ \textbf{ and } not converged} 

\State For each $i=1,\ldots, n$, set
\begin{equation*}
\theta_i \gets \beta^{k-1} + \sum_{j=1}^p f^{k-1}_j\left(x_{ij}\right), \quad\quad 
r_i \gets -\dot{\ell}\left(y_i,\theta_i\right).
\end{equation*}

\State Update 
$\beta^k\gets \beta^{k-1} - t\sum_{i=1}^{n}r_i.$

\For{$ j = 1,\, \ldots,\, p$} 
	\State Set 
	\begin{equation}\label{eq:update1}
	f^{inter}_j \gets \underset{f\in \mathcal{F}}{\operatorname{argmin }}\frac{1}{2} \left\|\left(f_j^{k-1} - t \bs{r}\right) - f\right\|_n^2 + t\lambda^2 P_{st}\left(f\right).
	\end{equation}
	\State Update
	\[
	f^{k}_j \gets \Big(1 - {t\lambda}/{\|f^{inter}_j\|_n}\Big)_{+}f^{inter}_j .
	\]
\EndFor
\EndWhile
\State\Return $\beta^k, f^k_1,\ldots, f^k_p$ 
\end{algorithmic}
\end{algorithm}

Algorithm \ref{alg:additive} is simple and can be quite fast: the time complexity is largely determined by the difficulty of solving the univariate smoothing problem of step $5$. In many cases this takes $O(n)$ operations, allowing an iteration of proximal gradient descent to run in $O(np)$ operations. Complexity order $O(np)$ is the per-iteration time complexity of state-of-the-art algorithms for the lasso \citep{friedman2010regularization,beck2009fast}.

Any step-size $t$ can be used in Algorithm \ref{alg:additive} so long as inequality~\eqref{eq:major} holds for $f_j^0 \equiv f_j^{k-1}$ and $f_j \equiv f_j^k$ when $(p+1)L$ is replaced by $t^{-1}$. Note that if $t \leq \{L(p+1)\}^{-1}$ this will always hold. However, often $p^k_{active}$, the number of $j$ for which either of $f^{k-1}_j$ or $f^k_j$ is non-zero, will be small. In this case $t \leq \{L(p_{active}+1)\}^{-1}$ will satisfy the majorization condition. Since, in practice, we are interested in sparse models, generally $p^k_{active} \ll p$ and adaptive step-size optimization can be quite useful \citep{beck2009gradient} . The algorithm can also take advantage of Nesterov-style acceleration \citep{nesterov2007gradient}, which improves the worst-case convergence rate after $k$ steps from $O\left({k}^{-1}\right)$ to $O\left({k^{-2}}\right)$. 

An important special case is the least squares loss $-\ell(y,\theta) = (y - \theta)^2$. In this case, we can use a block coordinate descent algorithm which can be more efficient than Algorithm~\ref{alg:additive}, and does not require a step-size calculation. We present the full details of the algorithm in Appendix~\ref{app:AdditionalFigures} in the supplementary material.
 
 As noted above, the main computational hurdle in Algorithm~\ref{alg:additive} is solving the univariate problem~\eqref{eq:opt2*}. In the following subsection, we discuss this step in greater detail for various smoothness penalties.

\subsection{Solving the Univariate Sub-problem}
For many semi-norm smoothers there are already efficient solvers for solving \eqref{eq:opt2*}: with the $k$-th order total variation penalty, \eqref{eq:opt2*} can be solved exactly in $2n$ operations for $k=0$ \citep{johnson2013dynamic}, or iteratively in roughly $O((k+1)n)$ operations for $k \geq 1$ \citep{ramdas2015fast}; with the convex indicator of an $M$-dimensional linear subspace, \eqref{eq:opt2*} can be solved in $O(M^2 n)$ operations using linear regression; using a monotonicity indicator, \eqref{eq:opt2*} can be solved with the pool adjacent violators algorithm in $O(n)$ operations \citep{ayer1955empirical}.

For many other choices of $P_{st}$, we do not have efficient algorithms for solving \eqref{eq:opt2*}; however, we might have fast algorithms for the slightly different optimization problem:
\begin{equation}\label{eq:spsq}
\wt{f}_{\wt{\lambda}} \gets \underset{f\in \mathcal{F}}{\operatorname{argmin }}\ \frac{1}{2} \left\| \bs{r} - f\right\|_n^2 + \wt{\lambda} P_{st}^{\tau}\left(f\right),
\end{equation}
for $\tau>1$. For example, the $k$-th order Sobolev penalty~\citep{wahba1990spline} can be solved exactly in $O(kn)$ operations for $\tau=2$. In the following Lemma, we show that the solution to \eqref{eq:spsq} can be leveraged to solve the harder problem \eqref{eq:opt2*}.

\begin{lemma}
\label{lemma:sqrtTrick}
Given an $n$-vector $\bs{r}$, a convex linear space $\mathcal{F}$ over the field $\mathbb{R}$, and real $\tau>1$, consider the optimization problems:
\begin{align*}
\wh{f}_{\lambda} &\gets \underset{f\in \mathcal{F}}{\operatorname{argmin }}\  \frac{1}{2}\left \| \bs{r} - f\right\|_n^2 + \lambda P_{st}\left(f\right);\label{eq:sobsat}\\
\wt{f}_{\lambda} &\gets \underset{f\in \mathcal{F}}{\operatorname{argmin }}\ \frac{1}{2}\left\| \bs{r} - f\right\|_n^2 + \lambda P_{st}^{\tau}\left(f\right);\\
f_{null} &\gets \underset{f\in \mathcal{F}}{\operatorname{argmin }}\ \frac{1}{2}\left\| \bs{r} - f\right\|_n^2 + \mathbb{I}\left(f \in \mathcal{F}: P_{st}(f)=0 \right);\\
f_{interp} &\gets \underset{f\in \mathcal{F}}{\operatorname{argmin }}\ P_{st}^{\tau}\left(f\right) + \mathbb{I}\left(r_i = f(x_i) \textrm{ for all $i$}\right),
\end{align*}
where $P_{st}(\cdot)$ is a semi-norm on $\mathcal{F}$. Assume that the directional derivative 
\begin{align*}
\nabla_{h} P^{\tau}_{st}(f) = \lim\limits_{\e\to 0} \frac{P^{\tau}_{st}(f+\e h) - P^{\tau}_{st}(f)}{\e},
\end{align*}
exists for all $h\in \mathcal{F}$. If $P_{st}(\wh{f}_{\lambda})\neq 0$ and $\tau\wt{\lambda}P^{\tau-1}_{st}(\wt{f}_{\wt{\lambda}}) = \lambda$, then $\wh{f}_{\lambda} = \wt{f}_{\wt{\lambda}}$.\\

\noindent To determine if $P_{st}(\wh{f}) = 0$, let $\mathcal{F} = \mathcal{F}_1\oplus \mathcal{F}_2$, where $\oplus$ is such that, for all $f\in \mathcal{F}$ we have $f = f_0+f_{\perp}$ where $\langle f_0,f_{\perp}\rangle_n = 0$ and $P_{st}(f) = P_{st}(f_{\perp})$. Furthermore, let $P_{st}^*$ be the dual norm over $\mathcal{F}_2$, given by 
\begin{equation}
\label{eqn:dual}
P_{st}^*(f_{\perp}) = \sup\  \ \Big\{ |\langle f_{\perp}, f_{\perp}'\rangle_n| : P_{st}(f_{\perp}') \le 1, f_{\perp}'\in \mathcal{F}_2 \Big\}.
\end{equation}
Then $f_{interp} - f_{null}\in \mathcal{F}_2$ and $\wh{f}_{\lambda} = f_{null}$ if $\lambda \ge P_{st}^*(f_{interp} - f_{null})$.
\end{lemma}
The proof is given in Appendix~\ref{sec:appendix} in the supplementary material. This lemma allows us to first check if we should shrink entirely to a null fit with $P_{st}(\wh{f}) = 0$ (usually a finite dimensional function), based on the dual semi-norm of the interpolating function $f_{interp}$. If we do not shrink to $P_{st}(\wh{f}) = 0$, then there is an equivalence between $\wh{f}$ and $\wt{f}$; and the problem is reduced to finding $\wt{\lambda}$ with $\tau\wt{\lambda}P_{st}^{\tau-1}(\wt{f}_{\wt{\lambda}}) = \lambda$ for the originally specified $\lambda$. This can be done in a number of ways; most simply by a combination of grid search and then local bisection noting that a) we need not try any $\wt{\lambda}$-values above $\lambda_{max} \equiv P_{st}\left(f_{interp}\right)$ (by Lemma~\ref{lemma:univ2}), and b) $\wt{\lambda}P_{st}(\wt{f}_{\wt{\lambda}})$ is a smooth function of $\wt{\lambda}$. In fact, the grid search will often be unnecessary as we will generally have a good guess from the previous iterate of the proximal gradient algorithm, and can leverage the fact that $P_{st}(\wt{f}_{\wt{\lambda}})$ and $P_{st}(\wh{f}_{\lambda})$ are both smooth functions of  $\bs{r}$.

To complete the discussion, we give the explicit form of the dual norm \eqref{eqn:dual} for the case where $P_{st}(f) = \|D\vec{f}\|_q$ for a matrix $D\in \mathbb{R}^{M\times n}$, a vector $\vec{f} = [f(x_1),\ldots, f(x_n)]^{\top}\in \mathbb{R}^n$, and  $q\ge 1$. Such penalties are common in the literature, e.g., when $P_{st}$ is the Sobolev semi-norm, total variation norm, or any RKHS norm. For $P_{st}(f) = \|D\vec{f}\|_q$, the dual norm is given by 
\begin{equation}
\label{eqn:dual2}
P_{st}^*(f) = \|D(D^{\top}D)^{-}\vec{f}\|_{\wt{q}},
\end{equation}
where $(D^{\top}D)^{-}$ is the Moore-Penrose pseudo inverse of $D^{\top}D$ and $\wt{q}$ satisfies $1/q + 1/\wt{q} = 1$.


\section{Theoretical Results}
\label{sec:TheoreticalResults}

Here we prove rates of convergence for \name s, estimators that fall within our framework~\eqref{eq:unif-sparse-add1}. We first present the so-called \emph{slow rates}, which require few assumptions, followed by \emph{fast rates}, which require compatibility and margin conditions (defined and discussed below). Our fast rates match the minimax rates under Gaussian data with a least squares loss~\citep{raskutti2009lower} and, our slow rates can be seen as an additive generalization of the lasso slow rates~\citep{dalalyan2017prediction}. For both slow and fast rates, we first present a deterministic result; 
this result simply states that if we are within a special set, $\mathcal{T}$, then the convergence rates hold. We then show that under suitable conditions (stated and discussed below) on the loss function, smoothness penalty, and data, we lie in $\mathcal{T}$ with high probability. Throughout, we also allow for mean model misspecification with an additional \emph{approximation error} term in the convergence rates; if the true mean model is additive, then this term disappears.

To the best of our knowledge, the closest results to our work were established by \cite{koltchinskii2010sparsity}. However, they consider a more restrictive setting of Reproducing Kernel Hilbert Spaces (RKHS); where each additive component $f_j$ belongs to a RKHS $\mathcal{H}_j$, and $P_{st}$ is the norm on $\mathcal{H}_j$. Our work gives these rates for all semi-norm penalties and function classes $\mathcal{F}$, associated with certain non-restrictive entropy conditions. Before presenting the main results, we present some notation and definitions which will be used throughout the section.
 
\subsection{Definitions and Notation}
We consider here properties of the solution to
\begin{equation}
\label{eq:prop}
\wh{\beta},  \wh{f}_1\ldots,\wh{f}_p \gets \underset{\beta\in \mathcal{R}, \{f_j\}_{j=1}^p\in \mathcal{F} }{\arg\min} -\mathbb{P}_n \ell\Big(\beta + \sum_{j=1}^p f_j\Big) + \lambda \sum_{j=1}^p\left\{\left\|f_j\right\|_n +  \lambda P_{st}\left(f_j\right)\right\} ,
\end{equation}
where $\mathcal{R}\subseteq \mathbb{R}$ and $\mathcal{F}$ is some univariate function class. Note that in \eqref{eq:prop} we optimize $\beta$ over $\mathcal{R}$; this is because we need $\mathcal{R}$ to be a bounded for proving the slow rates, the stronger \emph{compatibility condition} allows us to take $\mathcal{R} = \mathbb{R}$ for proving fast rates.

\noindent For a function $f(\bs{x}) = \beta + \sum_{j=1}^{p}f_j(x_j)$ we use the shorthand notation 
\begin{equation}
\label{eqn:defPenI}
I(f) \equiv \sum_{j=1}^p\left\{ \left\|f_j\right\|_n +  \lambda P_{st}\left(f_j\right)\right\},
\end{equation}
which defines a semi-norm on the function $f$. Furthermore, for any index set $S \subset \{1,\ldots, p\}$ we define $I_S(f)$ as 
$
I_S(f) = \sum_{j\in S}\left\{ \|f_j\|_n + \lambda P_{st}(f_j) \right\} .
$ We denote the target function by  $f^0$ where
\begin{equation}
\label{eqn:target}
f^0 \gets \underset{f\in \mathcal{F}^0}{\arg\min} -\mathbb{P} \ell\left(f\right),
\end{equation}
for some function class $\mathcal{F}^0$ and, where $\mathbb{P}\ell(f) = n^{-1}\sum_{i=1}^{n} \mathbb{E}\left\{\ell(y_i, f(\bs{x}_i))\right\}$. We say the target function belongs to some class $\mathcal{F}^0$ to signify that $f^0$ does not need to belong to $\mathcal{F}$. We require no assumptions on the class $\mathcal{F}^0$ for the slow-rates of Theorem~\ref{thm:main1}; we can take $\mathcal{F}^0$ to be the class of all measurable functions. For the fast rates we will require the margin condition on a subset of $\mathcal{F}^0$.

We define the \textit{excess risk} for a function $f$ as
$\mathcal{E}(f) = \mathbb{P}\left\{ \ell(f^0) - \ell(f) \right\},$
and we denote by $\nu_n(\cdot)$ the \textit{empirical process term}, which is defined as 
\begin{equation}
\label{eqn:empProcess}
\nu_n(f) = (\mathbb{P}_n - \mathbb{P})\left\{ -\ell(f) \right\} = -\frac{1}{n} \sum_{i=1}^{n} \left\{ \ell(y_i, f(\bs{x}_i))  - \mathbb{E}\ell(y_i, f(\bs{x}_i)) \right\}.
\end{equation}

Define the $\delta$-\textit{covering number}, $N(\delta, \mathcal{F}, \|\cdot\|_{Q})$, as the size of the smallest $\delta$-cover of $\mathcal{F}$ with respect to the norm $\|\cdot\|_Q$ induced by measure $Q$. We denote the $\delta$-\textit{entropy} of $\mathcal{F}$ by $H(\delta, \mathcal{F}, \|\cdot\|_Q) \equiv \log N(\delta,\mathcal{F}, \|\cdot\|_Q)$. Given fixed covariates $\bs{x}_1,\ldots,\bs{x}_n\in \mathbb{R}^p$, we denote the empirical measure by $Q_n$ where
$
Q_n = {n^{-1}}\sum_{i=1}^{n}\delta_{\bs{x}_i},
$
and for covariate $j$; we denote by $Q_{j,n}$ the empirical measure of $(x_{1,j},\ldots, x_{n,j})$. We define two different types of entropy bounds for a function class $\mathcal{F}$.

\begin{definition}[Logarithmic Entropy]
A univariate function class, $\mathcal{F}$, is said to have a logarithmic entropy bound if, for all $j=1,\ldots,p,$ and $\gamma>0$, we have 
\begin{equation}
	\label{eq:logEnt}
H(\delta, \left\{f_j\in  \mathcal{F}: \|f_j\|_n + \gamma P_{st}(f_j) \le 1 \right\}, \|\cdot\|_{Q_{j,n}}) \le A_0T_n\log\left( 1/\delta + 1\right),
\end{equation}
for some constant $A_0$, and parameter $T_n$.
\end{definition}

\begin{definition}[Polynomial Entropy with Smoothness]
	A univariate function class, $\mathcal{F}$, is said to have a polynomial entropy bound with smoothness if, for all $j=1,\ldots,p$ and $\gamma>0$, we have 
\begin{equation}
	\label{eq:polyEnt}
H(\delta, \left\{f_j\in  \mathcal{F}: \|f_j\|_n + \gamma P_{st}(f_j) \le 1 \right\}, \|\cdot\|_{Q_{j,n}}) \le A_0(\delta\gamma)^{-\alpha},
\end{equation}
for some constant $A_0$, parameter $\alpha\in (0,2)$.
\end{definition}
The concept of entropy is commonly used in the literature, particularly in nonparametric statistics and empirical processes, to quantify the size of function classes. The logarithmic entropy bound \eqref{eq:logEnt} holds for most finite dimensional classes of dimension $T_n$. For instance, it holds for $\mathcal{F} = L^2(\mathbb{R})$ with $P_{st}(f_j) = \mathbb{I}(f_j; \operatorname{span}\{x,x^2,\ldots, x^{T_n}\} )$. The bound \eqref{eq:polyEnt} commonly holds for broader function classes, e.g., for $\mathcal{F} = L^2([0,1])$ with $P_{st}(f_j) = P_{sobolev}(f^{(k)})$ and $\alpha = 1/k$.

\noindent To simplify our presentation of bounds on the convergence rate, we use $A \precsim B$ to denote $A\le cB$ for some constant $c>0$. We write $A\asymp B$ if $A\precsim B$ and $B\precsim A$.

\subsection{Main Results}
\label{sec:main}
We now present our main results: upper bounds for the excess risk of \name s, i.e., bounds for $\mathcal{E}(\wh{\beta} + \sum_{j=1}^{p}\wh{f}_j)$. The following theorem shows that $\mathcal{E}(\wh{\beta} + \sum_{j=1}^{p}\wh{f}_j) \precsim \lambda$ over a special set $\mathcal{T}$. In the corollary that follows, we show that for appropriate $\lambda$ values, and certain type of loss functions, we are within $\mathcal{T}$ with high probability.

\begin{theorem}[Slow Rates for \name]
\label{thm:main1}
Let $\wh{f} = \wh{\beta} + \sum_{j=1}^{p}\wh{f}_j$ be as defined in \eqref{eq:prop}, and let $f^* = \beta^* + \sum_{j=1}^{p}f_j^*$ be an arbitrary additive function with $\sum_{i=1}^{n}f^*_j(x_{ij}) = 0$ and $\beta^*\in \mathcal{R}$. Assume that $-\ell(\cdot)$ and $P_{st}$ are convex and that 
$
\sup_{\beta\in \mathcal{R}} |\beta| < R.
$
Define $M^*$ such that
\begin{equation}
\rho M^* = \mathcal{E}(f^*) + 2\lambda I(f^*) + 2R\rho,
\end{equation}
where 
$
\lambda \ge 4\rho. 
$
Furthermore, define the set $\mathcal{T}$ as follows
\begin{equation*}
\mathcal{T} = \left\{ Z_{M^*} \le \rho (M^* +2R) \right\}, \text{ where } Z_{M^*} = \underset{I(f - f^*)\le M^*}{\sup} \left|\nu_n(f) - \nu_n(f^*)\right|.
\end{equation*}
Then, on the set $\mathcal{T}$, 
\begin{equation*}
\mathcal{E}(\wh{f}) +  \lambda I(\wh{f}-f^*) \le \rho M^* + \rho (2R)+2\lambda I(f^*) + \mathcal{E}(f^*).
\end{equation*}
\end{theorem}

\begin{corollary}
\label{cor:slowRates}
Let $\wh{f},$ $f^*$ and $\mathcal{R}$ be as defined in Theorem~\ref{thm:main1}. Assume that for any function $f$ the loss $\ell(\cdot)$ is such that 
\begin{equation}
-\ell(f) = -\ell(y_i, f(\bs{x}_i)) = ay_if(\bs{x}_i) + b(f(\bs{x}_i)),
\end{equation}
for some $a\in \mathbb{R}\backslash \{0 \}$ and function $b:\mathbb{R}\to \mathbb{R}$. Further assume that for $i=1,\ldots,n$, $y_i - \mathbb{E}(y_i)$ are uniformly sub-Gaussian, i.e., 
\begin{equation}
\max_{i=1,\ldots,n} K^2\left[ \mathbb{E}\exp{\{y_i-\mathbb{E}(y_i)\}^2/K^2} - 1 \right] \le \sigma_0^2.
\end{equation}
Finally, suppose $\mathcal{E}(f^*) = O(\lambda)$ and $I(f^*) = O(1)$. Then, with probability at-least $1-2\exp\left( -C_1n\rho^2 \right)-C\exp\left( -C_2n\rho^2 \right) $, we have the following cases:
\newline

\begin{enumerate}
\item If $\mathcal{F}$ has a logarithmic entropy bound, then  for 
$\lambda \asymp \rho \asymp \kappa \max\left( \sqrt{\frac{T_n}{n}}, \sqrt{\frac{\log p}{n}} \right),$
\begin{equation}
\mathcal{E}(\wh{f}) + \lambda I(\wh{f} - f^*) \precsim \max\left( \sqrt{\frac{T_n}{n}}, \sqrt{\frac{\log p}{n}} \right),
\end{equation}
with constants $\kappa = \kappa(a, K,\sigma_0, A_0)$, $C_1 = C_1(K,\sigma_0)$, $C = C(K,\sigma_0)$ and $C_2 = C_2(C,\kappa)$.
\item If $\mathcal{F}$ has a polynomial entropy bound with smoothness, then for $\lambda \asymp \rho \asymp \kappa \max\left( n^{-\frac{1}{2+\alpha}}, \sqrt{\frac{\log p}{n}} \right),$ 
\begin{equation}
\mathcal{E}(\wh{f}) + \lambda I(\wh{f} - f^*)  \precsim \max\left( n^{-\frac{1}{2+\alpha}}, \sqrt{\frac{\log p}{n}} \right), 
\end{equation}
with constants $\kappa = \kappa(a, K,\sigma_0, A_0,\alpha)$, $C_1 = C_1(K,\sigma_0)$, $C = C(K,\sigma_0)$ and $C_2 = C_2(C,\kappa)$.\newline

\end{enumerate}
\end{corollary}

We now proceed to show the fast rates of convergence. To establish these rates, we require \textit{compatibility} and \textit{margin} conditions. The compatibility condition, is based on the idea that $I(f)$ and $\|f\|$ are somehow compatible for some norm $\|\cdot\|$. This condition is common in the high-dimensional literature for proving fast rates~(see \cite{geer2009conditions} for a discussion of compatibility and related conditions for the lasso). The margin condition, is based the idea that if $\mathcal{E}(f)$ is small then $\|f - f^0\|$ should also be small. This is another common condition in the literature for handling general convex loss functions~\citep[see e.g.,][]{negahban2011unified,geer2008high}. 

\begin{definition}[Compatibility Condition]
The compatibility condition is said to hold for an index set $S\subset \{1,2,\ldots, p\}$, with compatibility constant $\phi(S)>0$, if for all $\gamma>0$ and all functions $f$ of the form $f(\bs{x}) = \beta+\sum_{j=1}^{p}f_j(x_{j})$ that satisfy $\sum_{j\in S_*^c}\|f_j\|_n + \gamma \sum_{j=1}^{p}P_{st}(f_j) \le |\beta| + 3\sum_{j\in S_*} \|f_j\|_n$, it holds that
\begin{equation}
\label{eqn:compatibility}
|\beta|/2 + \sum_{j\in S_*}\|f_j\|_n \le \|f\|\sqrt{|S|}/\phi(S),
\end{equation}
for some norm $\|\cdot\|$.
\end{definition}

\begin{definition}[Margin Condition]
The margin condition holds if there is strictly convex function $G$ such that $G(0) = 0$ and for all $f\in \mathcal{F}^0_{local} \subset \mathcal{F}^0$ we have 
\begin{equation}
\mathcal{E}(f) \ge G(\|f - f^0\|),
\end{equation}
for some norm on the function class $\mathcal{F}^0$; here $\mathcal{F}^0_{local}$ is a neighborhood of $f^0$ based on some norm~(e.g., $\mathcal{F}^0_{local} = \{f: \|f-f^0\|_{\infty} \le \eta \}$). In typical cases, the margin condition holds with $G(u) = cu^2$, for a positive constant $c$. We refer to this special case as the quadratic margin condition.
\end{definition}

The following theorem establishes the bound $\mathcal{E}(\wh{\beta} + \sum_{j=1}^{p}\wh{f}_j) \precsim s\lambda^2$, where $\lambda$ is the slow rate of Theorem~\ref{thm:main1}, and $s$ is the number of non-zero components of $f^* = \beta + \sum_{j=1}^{p} f^*_j$, a sparse additive approximation of $f^0$. As in Theorem~\ref{thm:main1}, the bound holds over a set $\mathcal{T}$; Corollary~~\ref{cor:fastrates} following the theorem shows that we lie in $\mathcal{T}$ with high probability.

\begin{theorem}[Fast Rates for \name]
\label{thm:main2a}
Suppose $-\ell\left(\cdot\right)$ and $P_{st}$ are convex functions and with $\wh{f}$ and $f^*$ as defined in Theorem~\ref{thm:main1}. Assume that $f^*$ is sparse with $|S_*| = s$ where 
$
S_* = \{j: f^*_j\not= 0\},
$
and that the compatibility condition holds for $S_*$. Further assume the quadratic margin condition holds with constant $c$, and that for a function $f(\bs{x}) = \beta +\sum_{j=1}^{p}f_j(x_j)$, $f\in \mathcal{F}^0_{local}$ if and only if
$|\beta - \beta^*| + I(f - f^*) \le M^*.$
The constant $M^*$ is defined as 
$$
\rho M^* = \mathcal{E}(f^*) + \frac{16s\lambda^2}{c\phi^2(S_*)}  + 2\lambda^2 \sum_{j\in S_*}P_{st}(f_j^*) ,
$$
and $\rho$ is such that 
$
\lambda \ge 8\rho.
$
Furthermore, define the set $\mathcal{T}$ as
\begin{equation*}
\mathcal{T} = \left\{ Z_{M^*} \le \rho M^* \right\}, \text{ where } Z_{M^*} = \underset{|\beta-\beta^*| + I(f - f^*) \le M^*}{\sup} \left|\nu_n(f) - \nu_n(f^*)\right|.
\end{equation*}
Then, on the set $\mathcal{T}$, 
\begin{align}
\mathcal{E}(\wh{f}) + \lambda I(\wh{f} - f^*) \le 4\rho M^* = 4\mathcal{E}(f^*) + \frac{64s\lambda^2}{c\phi^2(S_*)} + 8\lambda^2 \sum_{j\in S_*} P_{st}(f_j^*).
\end{align}
\end{theorem}

\begin{corollary}
\label{cor:fastrates}
Let $\wh{f}$ and $f^*$ be as defined in Theorem~\ref{thm:main1} and assume the conditions of Theorem~\ref{thm:main2a}. Furthermore, for any function $f$ assume the loss $\ell(\cdot)$ is such that 
\begin{equation}
-\ell(f) = -\ell(y_i, f(\bs{x}_i)) = ay_if(\bs{x}_i) + b(f(\bs{x}_i)),
\end{equation}
for some $a\in \mathbb{R}\backslash \{0 \}$ and function $b:\mathbb{R}\to \mathbb{R}$. Further assume that for $i=1,\ldots, n$, $y_i - \mathbb{E}y_i$ are uniformly sub-Gaussian, i.e. 
\begin{equation}
\max_{i=1,\ldots,n} K^2\left[ \mathbb{E}\exp\left\{ (y_i-\mathbb{E}y_i)^2/K^2 \right\} - 1 \right] \le \sigma_0^2.
\end{equation}
Finally suppose $\mathcal{E}(f^*) = O(s\lambda^2/\phi^2(S_*))$ and $s^{-1}\sum_{j\in S_*}P_{st}(f_j^*) = O(1)$. Then, with probability at-least $1-2\exp\left( -C_1n\rho^2 \right)-C\exp\left( -C_2n\rho^2 \right)$, we have the following cases:
\newline
\begin{enumerate}
\item If $\mathcal{F}$ has a logarithmic entropy bound, for 
$\lambda \asymp \rho \asymp \kappa \max\left( \sqrt{\frac{T_n}{n}}, \sqrt{\frac{\log p}{n}} \right)$,
\begin{equation}
\mathcal{E}(\wh{f}) + \lambda I(\wh{f} - f^*) \precsim \max\left( s{\frac{T_n}{n}}, s{\frac{\log p}{n}} \right),
\end{equation}
with constants $\kappa = \kappa(a, K,\sigma_0, A_0)$, $C_1 = C_1(K,\sigma_0)$, $C = C(K,\sigma_0)$ and $C_2 = C_2(C,\kappa)$.


\item If $\mathcal{F}$ has a polynomial entropy bound with smoothness, then for $\lambda \asymp \rho \asymp \kappa \max\left( n^{-\frac{1}{2+\alpha}}, \sqrt{\frac{\log p}{n}} \right),$
\begin{equation}
\label{eqn:fastRate}
\mathcal{E}(\wh{f}) + \lambda I(\wh{f} - f^*)  \precsim \max\left( sn^{-\frac{2}{2+\alpha}}, s{\frac{\log p}{n}} \right),
\end{equation}
with constants $\kappa = \kappa(a, K,\sigma_0, A_0,\alpha)$, $C_1 = C_1(K,\sigma_0)$, $C = C(K,\sigma_0)$ and $C_2 = C_2(C,\kappa)$.\newline
\end{enumerate}
\end{corollary}

We will discuss the significance of our theoretical results in the next subsection by specializing them to some well-studied special cases. Before discussing these specializations, we conclude this section by further generalizing   Theorem~\ref{thm:main2a}. We will now assume a more general margin condition,  for which we need to define the additional notion of a \textit{convex conjugate}.

\begin{definition}[Convex Conjugate]
Let $G$ be a strictly convex function on $[0,\infty)$ with $G(0) = 0$. The convex conjugate of $G$, denoted by $H$, is defined as
\begin{equation}
H(v) = \sup_{u} \left\{ uv - G(u) \right\}, \ v\ge 0.
\end{equation}
For the special case of $G(u) = cu^2$, one has $H(v) = v^2/(4c)$.
\end{definition}

\begin{theorem}[Fast Rates]
\label{thm:main2}
Assume the conditions of Theorem~\ref{thm:main2a} and define $M^*$ as
\begin{equation}
\rho M^* = \mathcal{E}(f^*) + H\left( \frac{8\lambda\sqrt{s}}{\phi(S_*)} \right) + 2\lambda^2 \sum_{j\in S_*}P_{st}(f_j^*),
\end{equation}
where $H(\cdot)$ is the convex conjugate of $G$. Then, on the set $\mathcal{T}$, 
\begin{align}
\mathcal{E}(\wh{f}) + \lambda I(\wh{f} - f^*) \le 4\rho M^*.
\end{align}
\end{theorem}

Note that our convergence rates include the term $\sum_{j\in S_*} P_{st}(f_j^*)$, or constants which depend on it. For some choices of $P_{st}$ this can lead to poor finite sample performance.  In such cases, prediction performance can be improved by solving instead
\begin{equation} \label{eq:unif-sparse-addAlternative}
\small
\wh{\beta}, \wh{f}_1,\ldots,\wh{f}_p \gets \hspace{-3mm} \underset{\beta\in \mathbb{R},f_1,\ldots,f_p \in \mathcal{F}}{\operatorname{argmin}}  -\mathbb{P}_n \ell\left(\beta + \sum_{j=1}^p f_j\right) + \omega\lambda^2\sum_{j=1}^p P_{st}\left(f_j\right) +  (1-\omega)\lambda\sum_{j=1}^p\left\|f_j\right\|_n,
\end{equation} 
where $\omega\in [0,1]$ is an additional tuning parameter. In Section~\ref{sec:SimulationStudy}, we empirically observe that the single tuning parameter formulation~\eqref{eq:unif-sparse-add1} is adequate for various choices of smoothness norms.

\noindent\textbf{Note on convex indicator penalties:} The above results do not directly extend to some convex indicator penalties. For some convex indicator penalties, such as $P_{st}(f) = \mathbb{I}(f;  \{f: f^{\prime} \ge 0 \})$, we require a third type of entropy condition: 
\begin{definition}[Polynomial Entropy without Smoothness]
The univariate function class, $\mathcal{F}$, is said to have a polynomial entropy without smoothness bound if for all $j=1,\ldots,p$ we have 
\begin{equation}
H(\delta, \left\{f_j\in  \mathcal{F}: \|f_j\|_n + \gamma P_{st}(f_j) \le 1 \right\}, \|\cdot\|_{Q_{j,n}}) \le A_0\delta^{-\alpha},
\end{equation}
for some constant $A_0$, parameter $\alpha\in (0,2)$ and all $\gamma>0$.
\end{definition}
Our results do not extend to convex indicator penalties because our proof relies on the fact that $f_j - f_j^* \in \mathcal{F}$ for $f_j,f_j^*\in \mathcal{F}$; function classes with polynomial entropy without smoothness do not usually have this property. We defer the extension to convex indicator structural penalties to future work.

\subsection{Special Cases of \name}
In this subsection, we illustrate the main strength of our framework, namely its generalizability. We specialize our theoretical results to, various existing proposals for sparse additive models, low-dimensional additive models, and fully non-parametric regression problems. We also specialize our results to GLMs in low and high dimensions.

As discussed in Section~\ref{sec:prevWork}, \cite{meier2009high} proposed a \name\ with $P_{st}(f) = P_{sobolev}(f^{\prime\prime})$. However, in their theoretical analysis they considered a larger class of structural penalties, namely penalties which satisfy the \emph{polynomial entropy with smoothness} condition \eqref{eq:polyEnt}. \cite{meier2009high} establish a convergence rate of the order $s(\log p/n)^{2/(2+\alpha)}$ which is sub-optimal compared to our fast rate \eqref{eqn:fastRate}. Established rates for the diagnolized smoothness penalty of \cite{geer2010lasso}, were also sub-optimal and of the order $s(\log p)n^{-2/(2+\alpha)}$. Our work bridges the following gaps in the theoretical work of \cite{meier2009high} and \cite{geer2010lasso}: (a) we establish minimax rates under identical compatibility conditions, (b) we extend their result beyond least squares loss functions and, (c) we establish \emph{slow rates} under virtually no assumptions. 

Another special case is trend filtering additive models \citep{petersen2016fused,sadhanala2017additive}. Theorem~\ref{thm:main1} improves upon the slow rates established by \cite{petersen2016fused} of the order $\sqrt{\log(np)/n}$; Theorem~\ref{thm:main2a} establishes fast rates by solving the problem which \cite{sadhanala2017additive} characterized as ``... still an open problem''. 

Additive models in low dimensions can also be considered by simply setting $S_* = \{1,\ldots,p\}$. In this case, the compatibility condition holds and we recover the usual convergence rates for generalized additive models of the form $p n^{-{2}/(2+\alpha)}$. With this, we recover the special case of univariate nonparametric regression, i.e., with $p=1$. Another interesting case that we recover is the multivariate nonparametric regression problem; to see this, suppose we have a single (but multivariate) component function $f_1:\mathbb{R}^p\to \mathbb{R}$. For various choices of $P_{st}$, the bound \eqref{eq:polyEnt} holds with $\alpha = p/m$ for some smoothness parameter $m$. Thus, we recover the usual nonparametric rate $n^{-2m/(2m+p)}$. 

Finally, parametric regression models are also special cases of \name. Using a convex indicator for $P_{st}$, we can constrain each $f_j$ to be a linear function leading to GLMs. For low-dimensional GLMs, Corollary~\ref{cor:fastrates} gives the usual parametric rate, $p/n$. For high-dimensional GLMs, not only does our theorem recover the lasso rate, but our compatibility condition also matches that of lasso~\citep{buehlmann2011statistics}. 


\begin{figure}
\centering
\includegraphics[scale = 0.25]{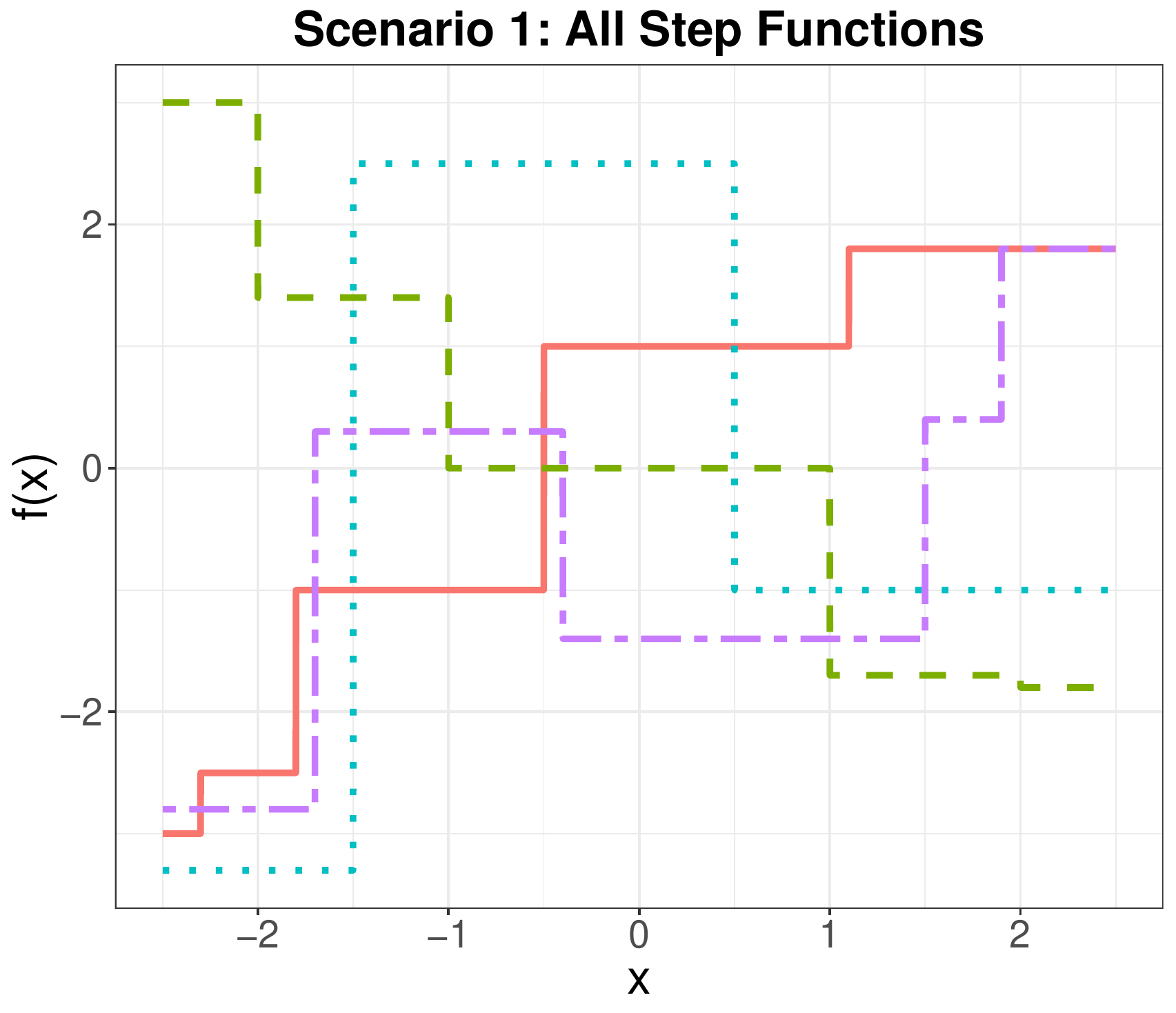}
\includegraphics[scale = 0.25]{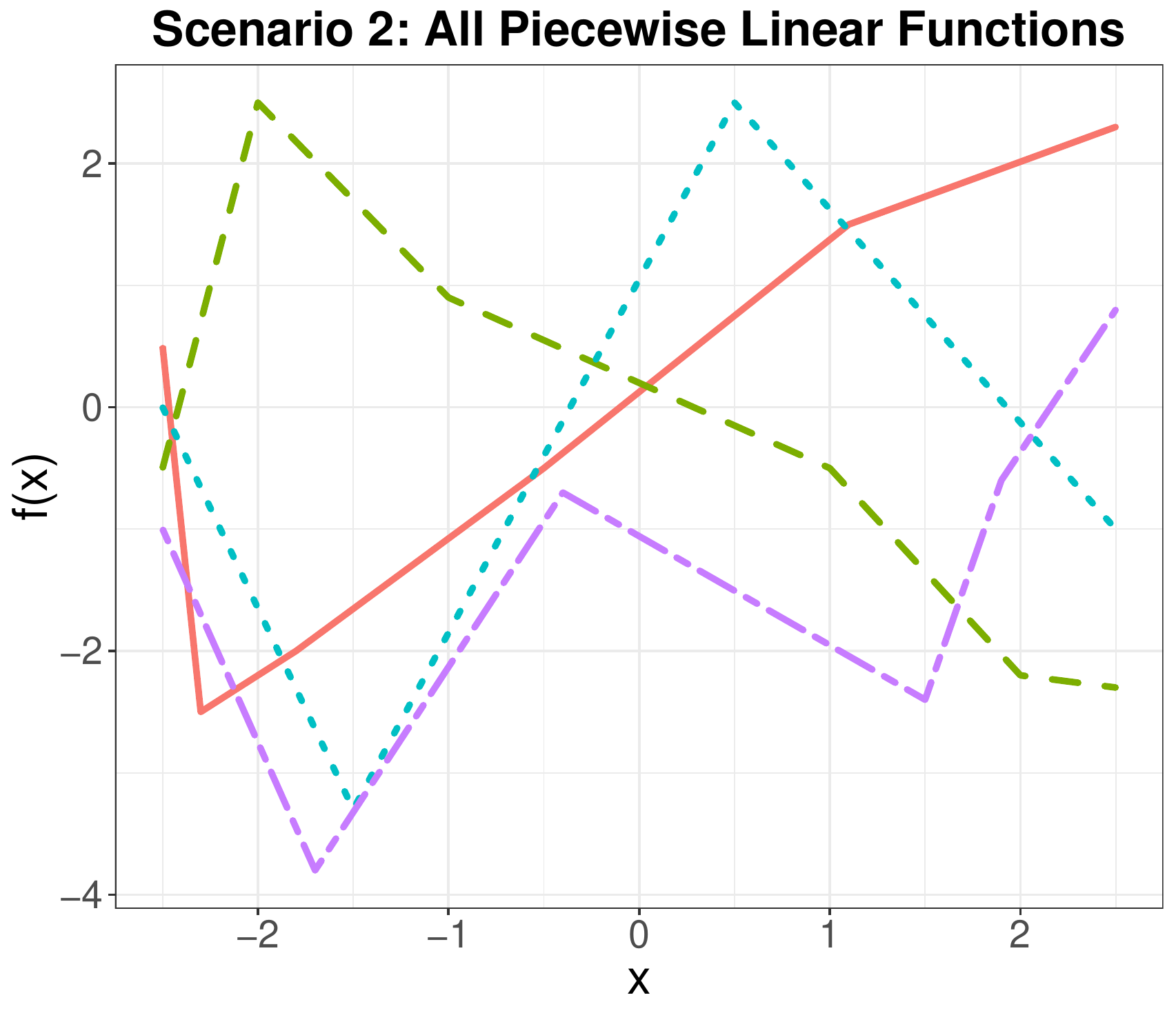}
\includegraphics[scale = 0.25]{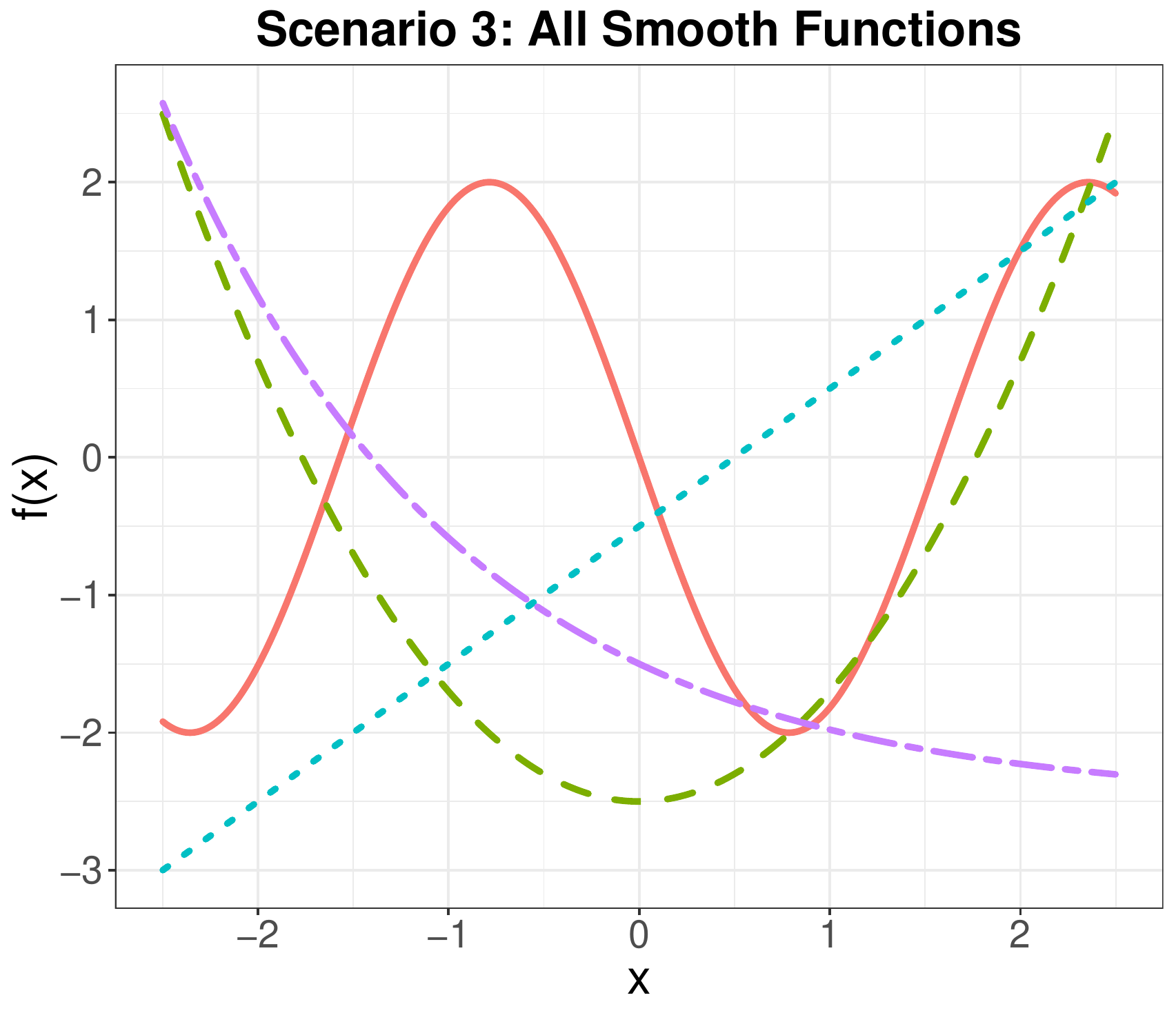}
\includegraphics[scale = 0.25]{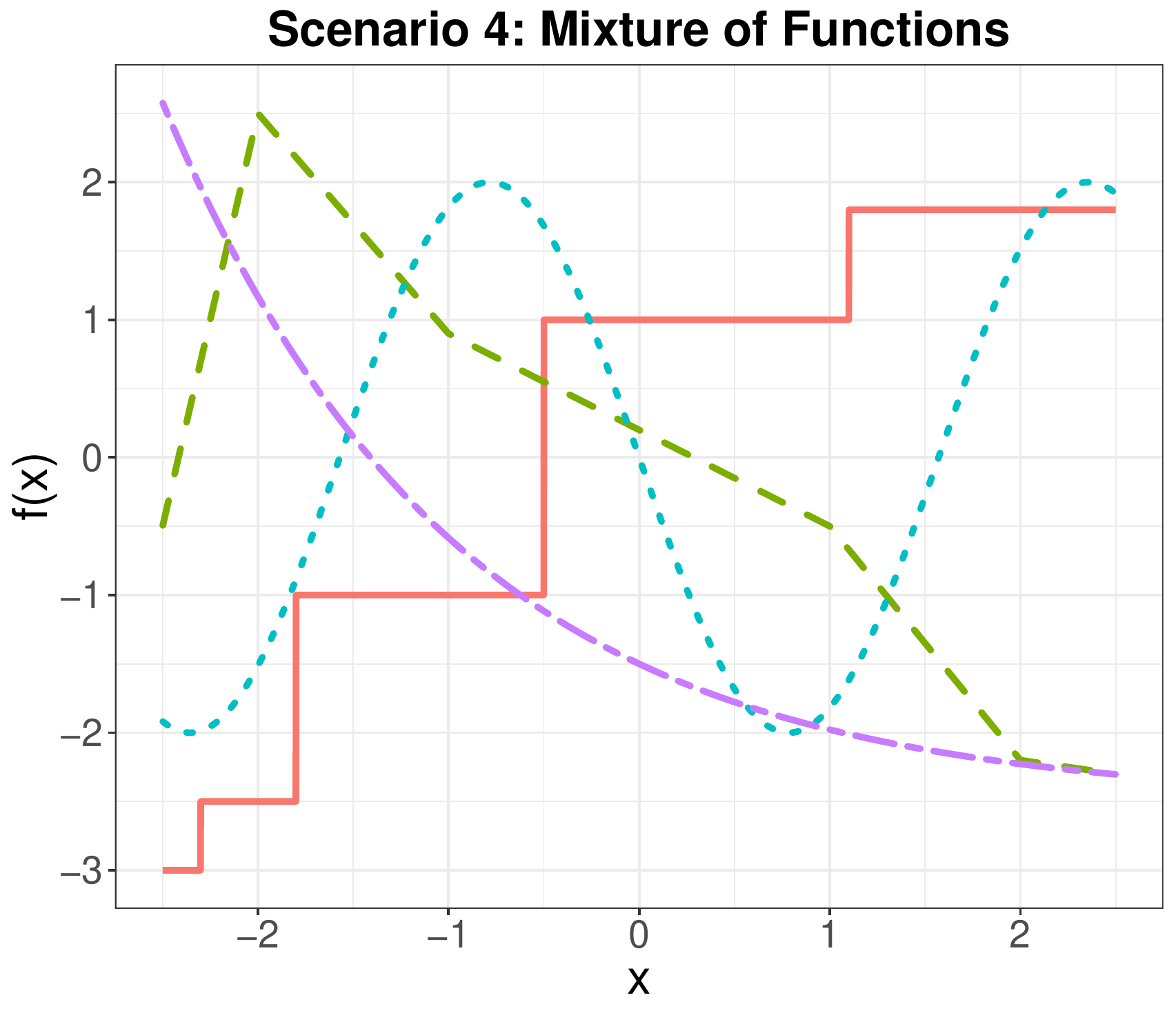}
\includegraphics[scale = 0.25]{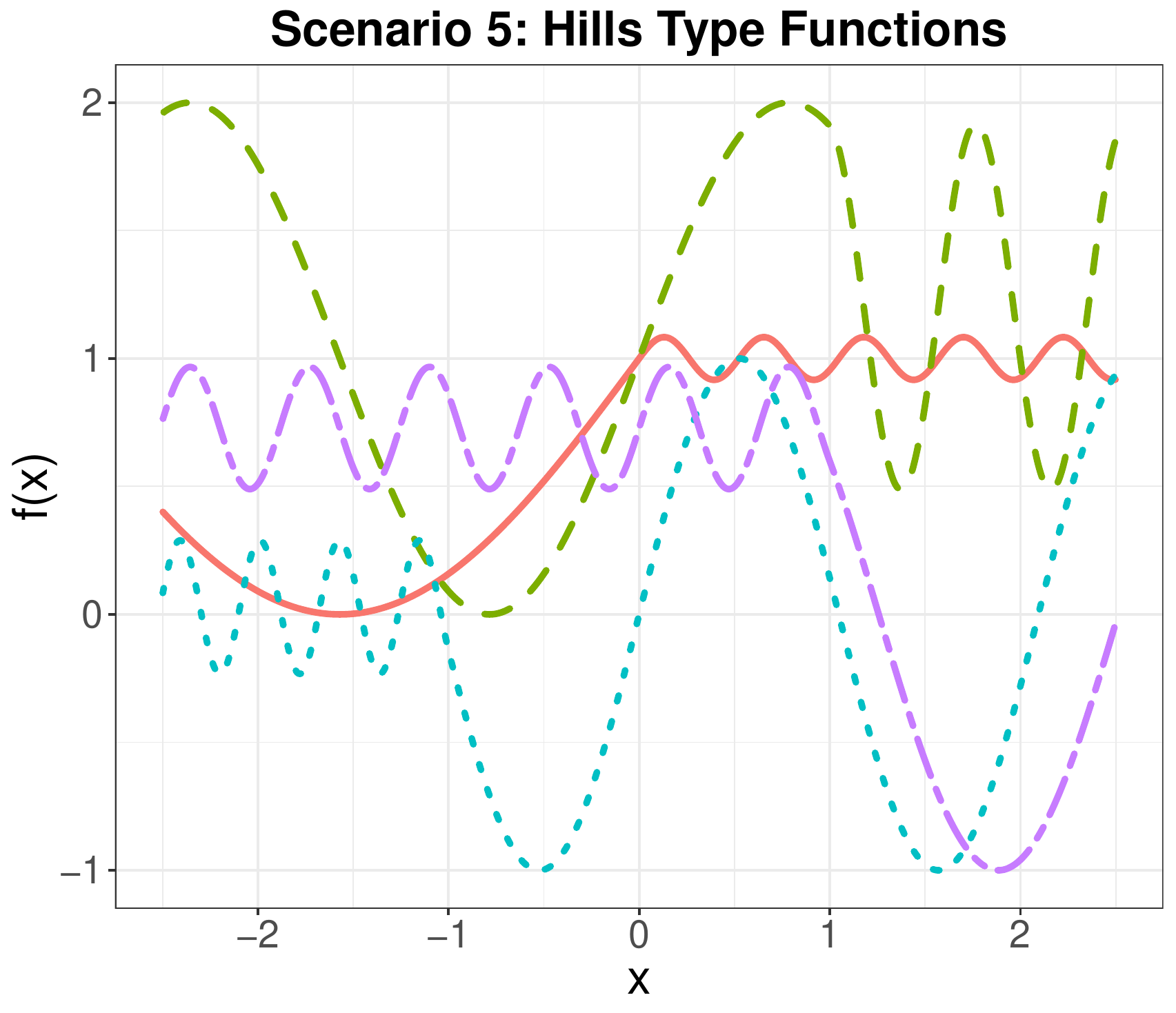}
\caption{Plot of the 4 signal functions for each of the five simulation settings.}
\label{fig:models}
\end{figure}
\section{Simulation Study}
\label{sec:SimulationStudy}
In this section we conduct a simulation study to compare estimators obtained by the following choices of smoothness penalty, $P_{st}(\cdot)$.
\begin{enumerate}
\item \textbf{SpAM~\citep{ravikumar2009sparse}.} $P_{st}(f) = \mathbb{I}(f; \operatorname{span}\{ \psi_1,\ldots,\psi_M \})$ for $M \in \{3,6,10,20,30,50,80\}$. We use the \texttt{SAM} \texttt{R}-package \citep{zhao2014SAM}.
\item \textbf{SSP~\citep{meier2009high}.} $P_{st}(f) = \sqrt{\int_x ( f^{(2)}(x))^2\, dx}$, the Sobolev smoothness penalty~(SSP). Given the lack of efficient software for this method, we implemented it using the algorithm and results of Section~\ref{sec:Algorithm}.
\item \textbf{TF~\citep{sadhanala2017additive}.} $P_{st}(f) = \int_x |f^{(k+1)}(x)|\, dx$ for $k\in \{0,1,2\}$, trend filtering for additive models. We implemented this method using the algorithm of Section~\ref{sec:Algorithm} where the univariate sub-problem~\eqref{eq:update1} was solved using the \texttt{R} package \texttt{glmgen}~\citep{arnold2014glmgen}.
\end{enumerate}

We simulate data for each of five simulation scenarios as follows: Given a sample size  $n$ and a number of covariates $p$, we draw 50 different $n\times p$ training design matrices $X$ where each element is drawn from $\mathcal{U}(-2.5,2.5)$. We replicate each of the 50 design matrices 10 times leading to a total of 500 design matrices. The response is generated as $y_i = f_1(x_{i1}) + f_2(x_{i2}) +f_3(x_{i3}) +f_4(x_{i4})+ \e_i$ where $\e_i\sim \mathcal{N}(0,1)$. The remaining covariates are noise variables. We also generate an independent test set for each replicate with sample size $n/2$. We vary the sample size, $n\in \{100,200, \ldots, 800\}$ and consider both, a low-dimensional ($p=6$) and high-dimensional~($p=100$) settings. We consider five different choices of the signal functions as shown in Figure~\ref{fig:models}.

We fit each method over a sequence of $50~\lambda$ values on the training set, and select the tuning parameter $\lambda^*$ which minimizes the test error~($\|y_{test}-\wh{y}\|^2_n$). For the estimated model $\wh{f}_{\lambda^*}$, we report the mean square error~(MSE; $\|\wh{f}_{\lambda^*} - f^0\|_n^2$) as a function of $n$. 

\begin{figure}
	\centering
	\includegraphics[width=0.31\textwidth]{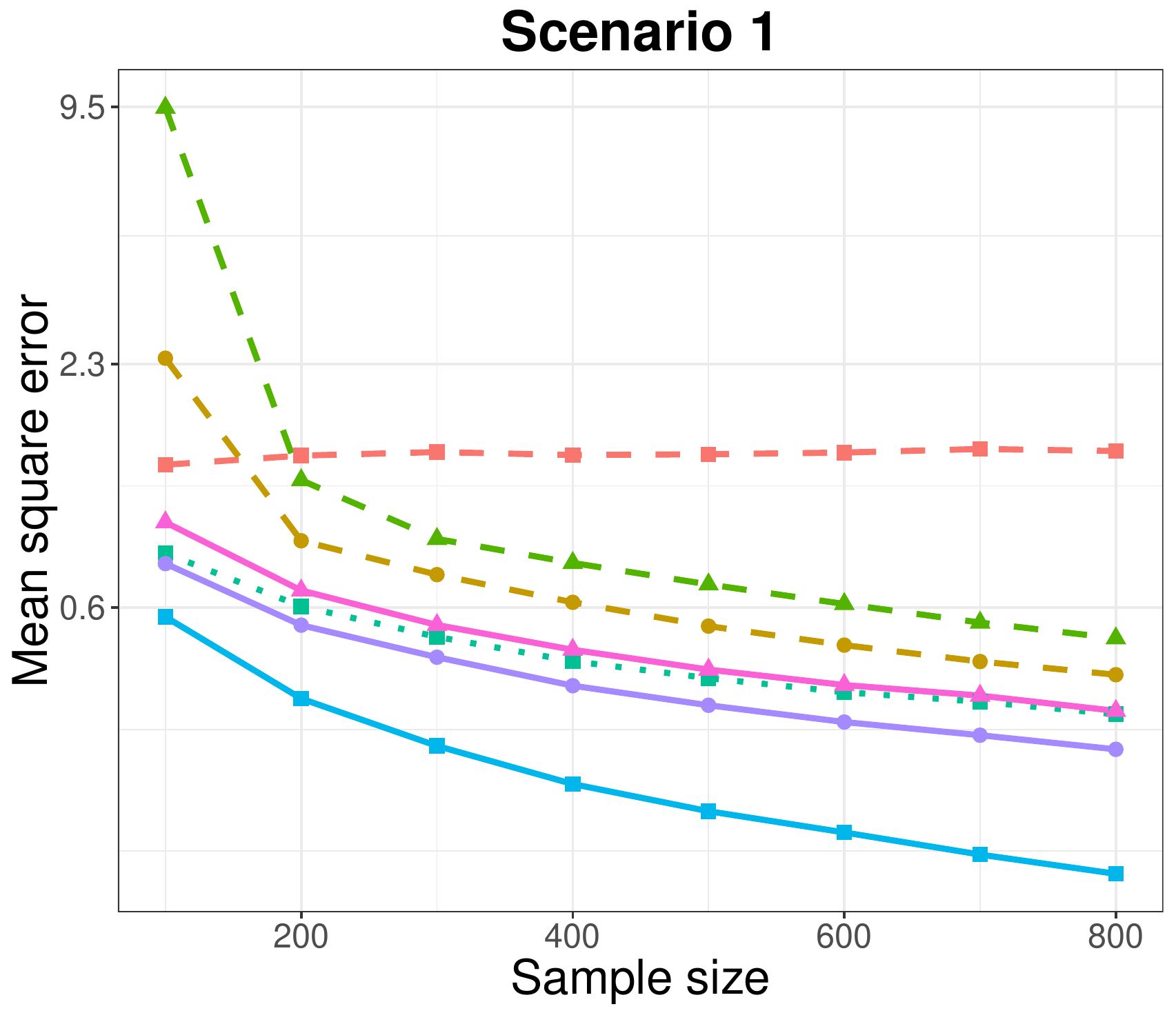}
	\includegraphics[width=0.31\textwidth]{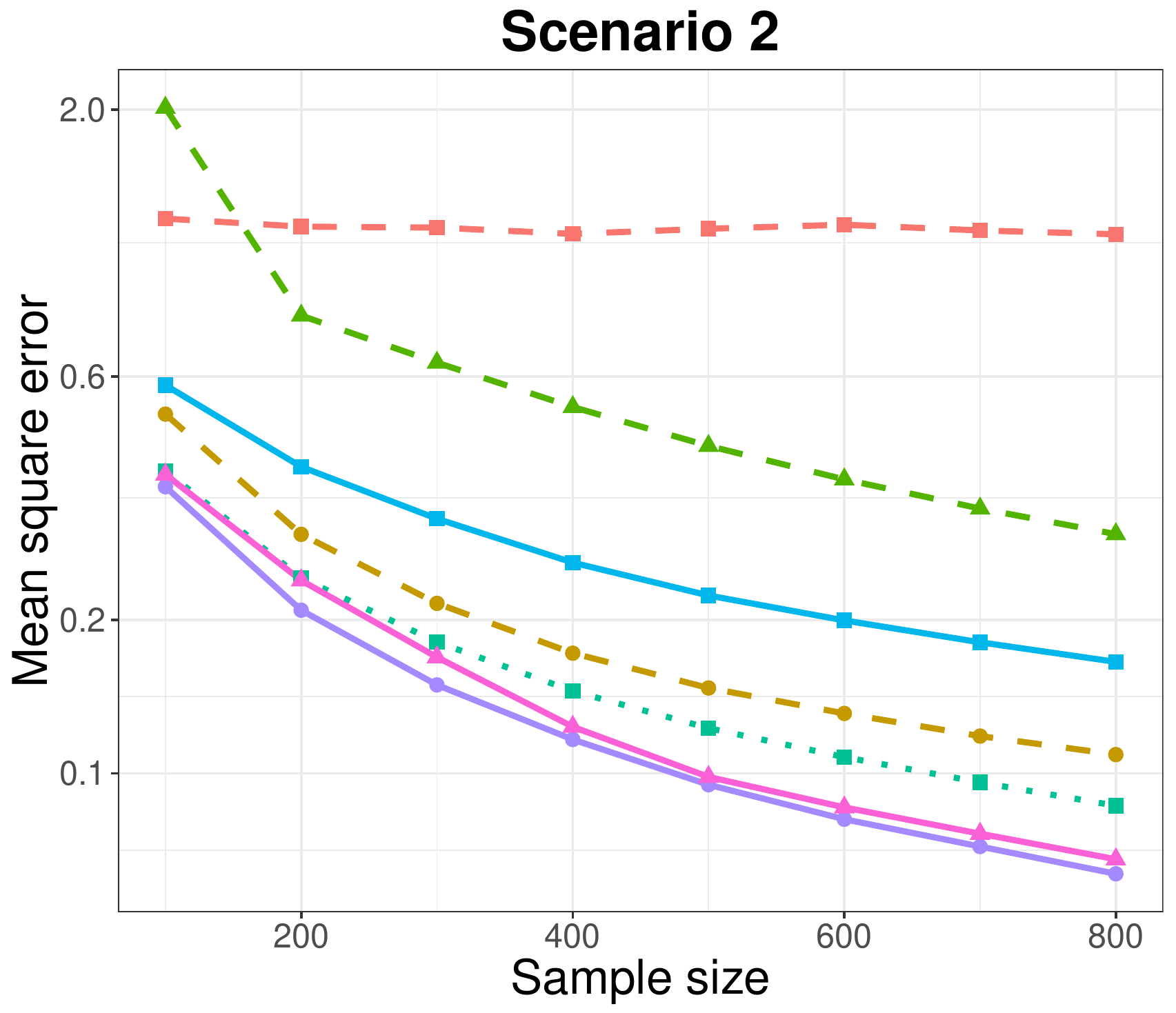}
	\includegraphics[width=0.31\textwidth]{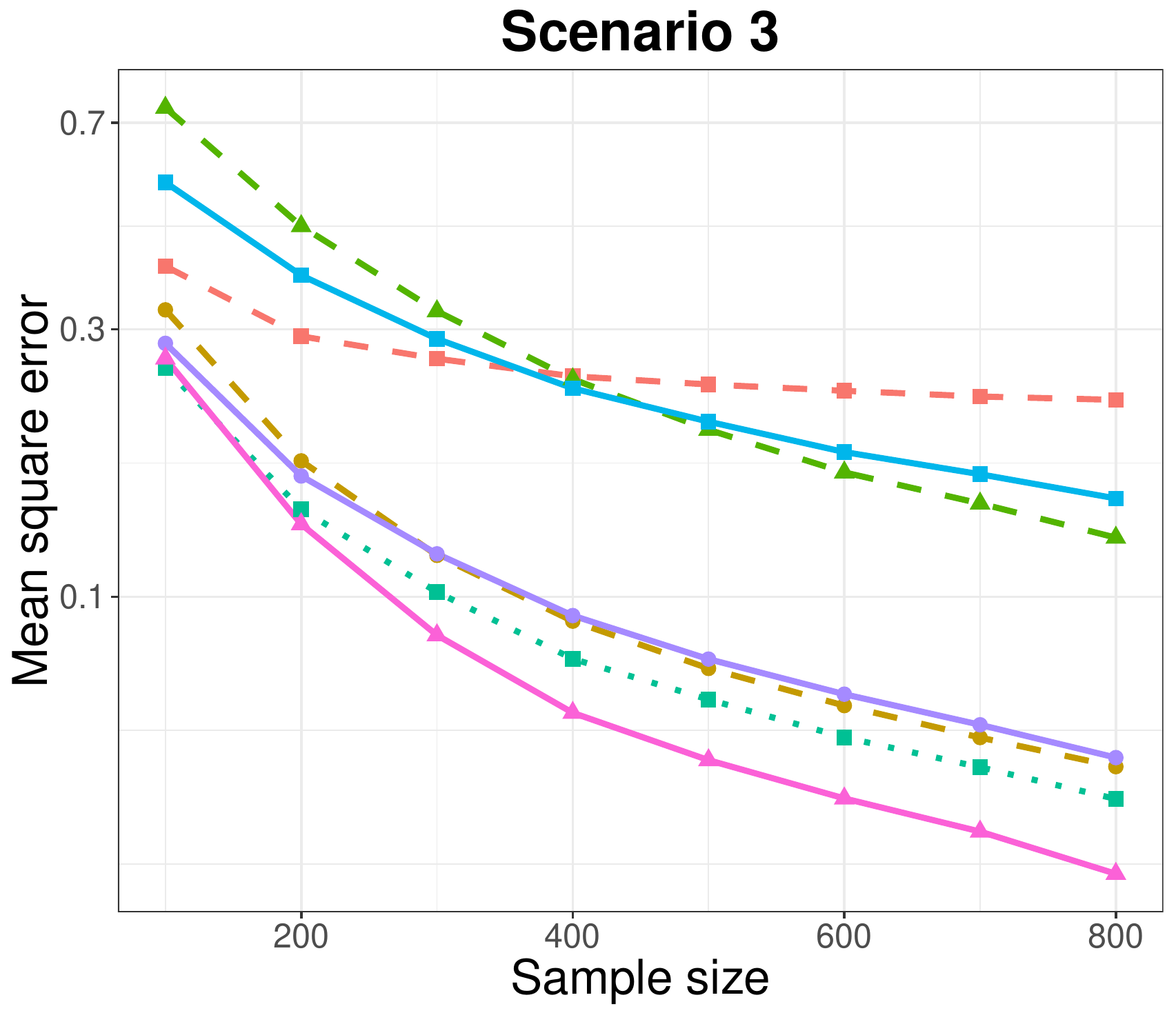}
	\includegraphics[width=0.31\textwidth]{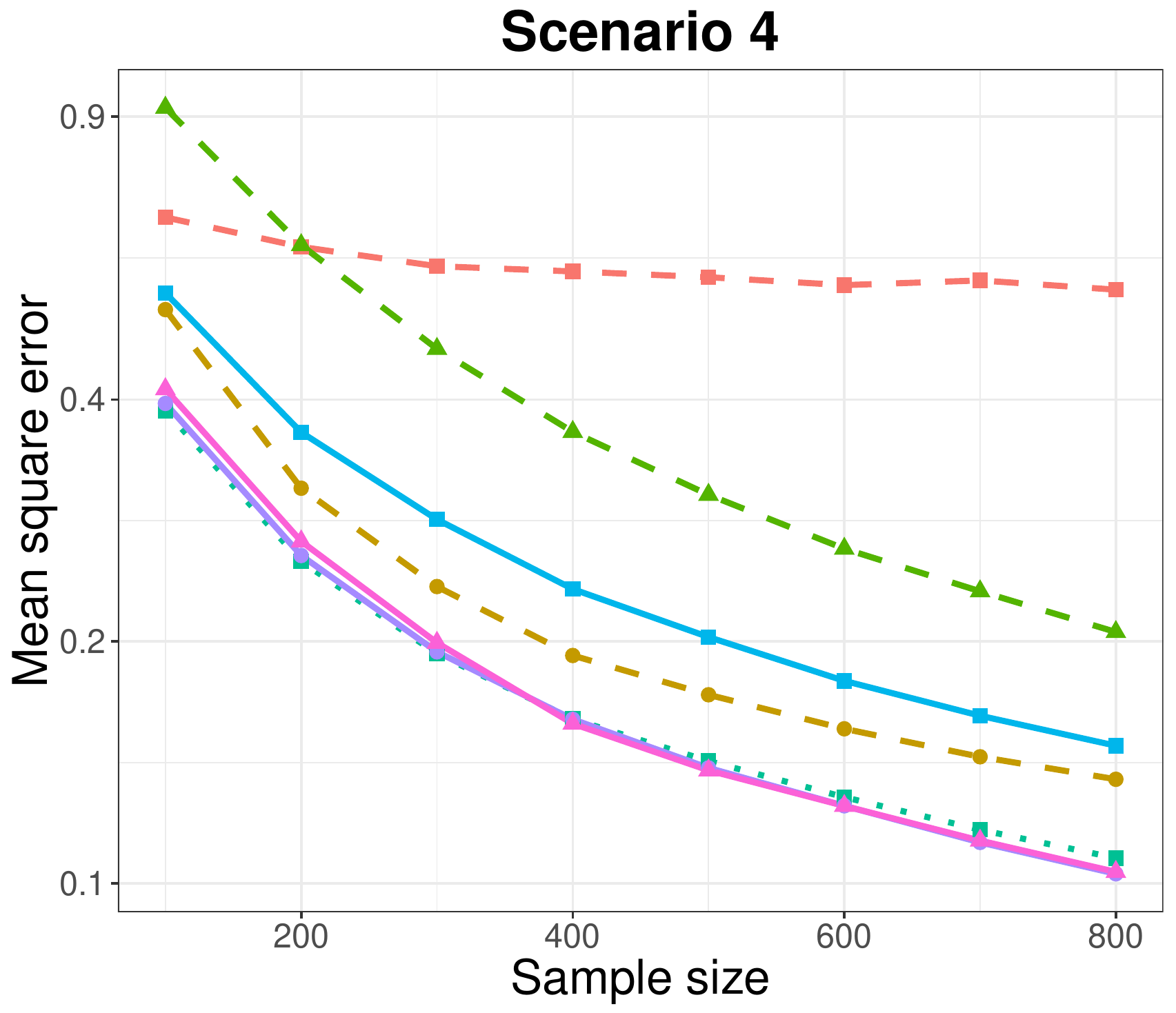}
	\includegraphics[width=0.31\textwidth]{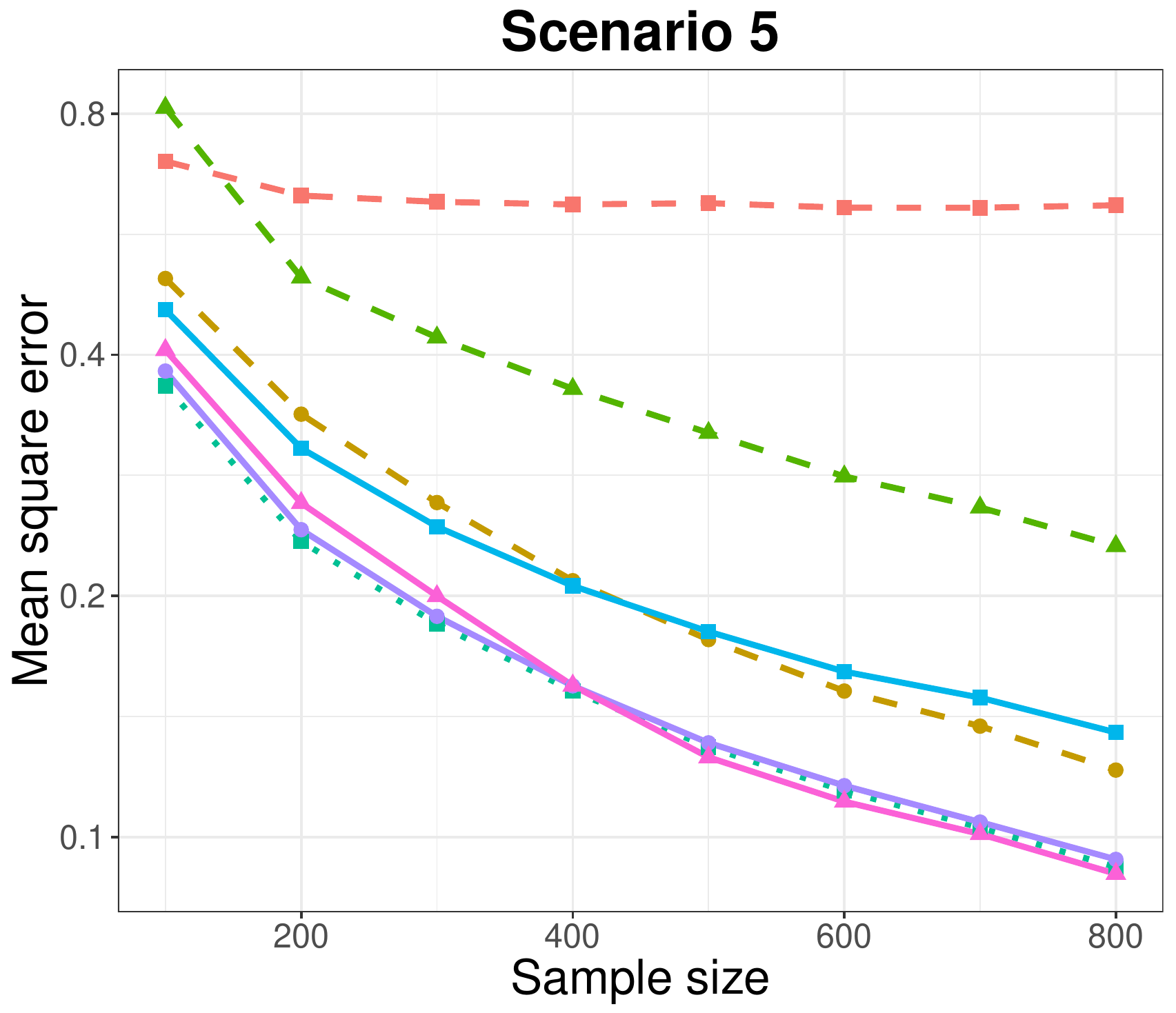}
	\caption{Plot of MSEs versus sample size for each of five scenarios for $p=6$, averaged over 500 replications. The dashed lines correspond to \texttt{SpAM} with small~(\protect\includegraphics[height=0.5em]{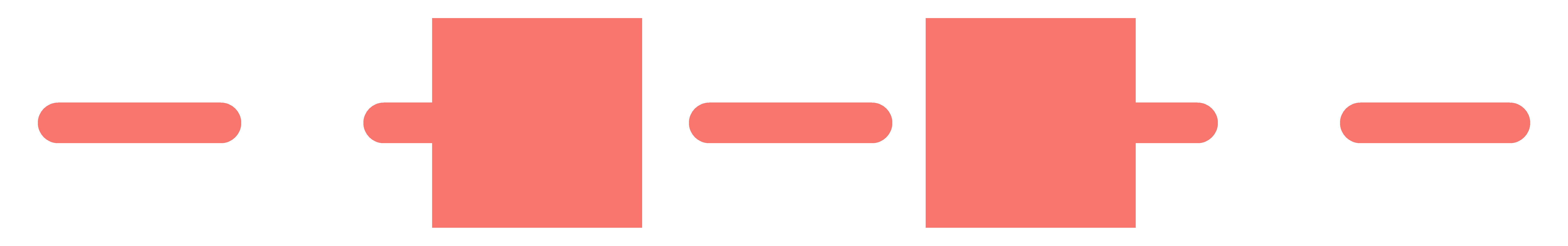}), moderate~(\protect\includegraphics[height=0.5em]{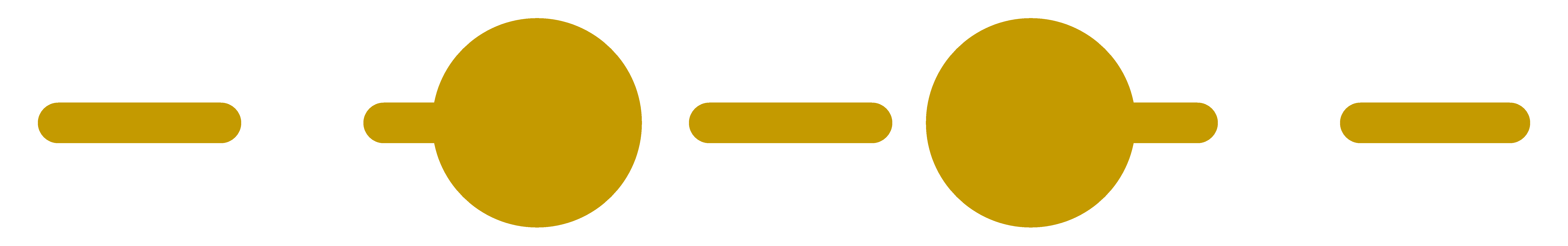}) and high~(\protect\includegraphics[height=0.5em]{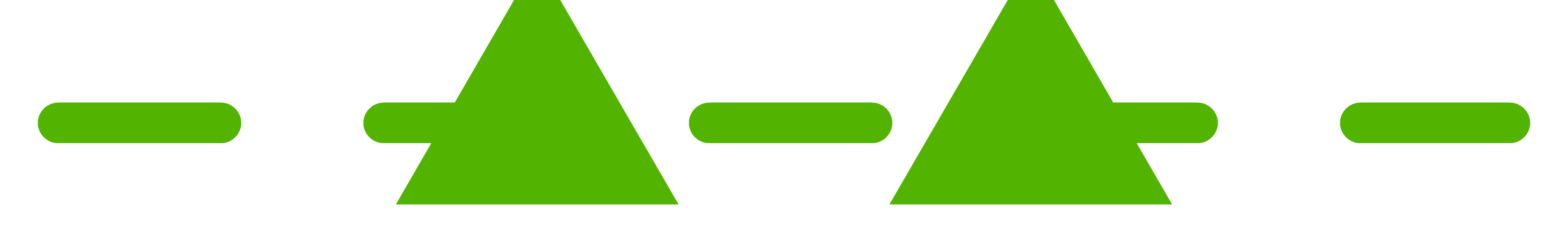}) number of basis functions. The solid lines correspond to trend filtering of order $k=0$~(\protect\includegraphics[height=0.5em]{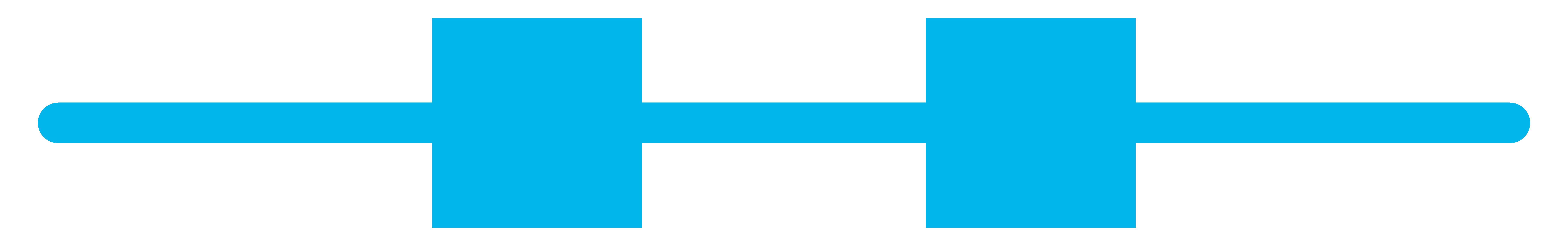}), 1~(\protect\includegraphics[height=0.5em]{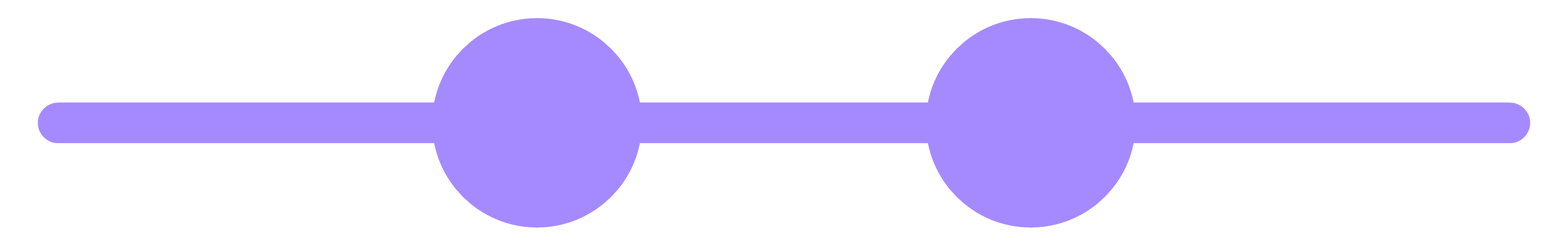}) and 2~(\protect\includegraphics[height=0.5em]{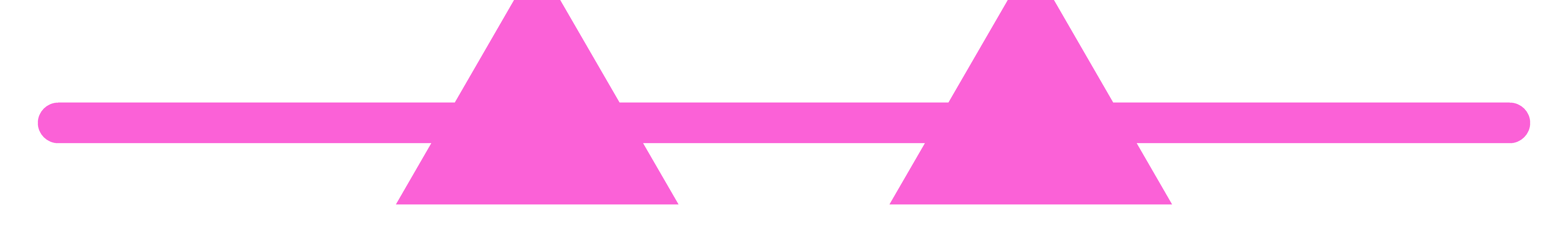}). SSP is represented by the dotted line~(\protect\includegraphics[height=0.5em]{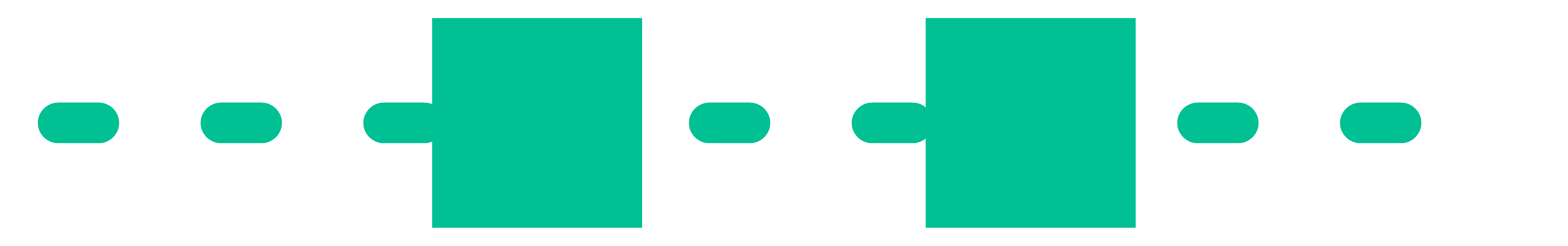}).}
	\label{fig:MSEp6}
\end{figure}

\begin{figure}
	\centering
	\includegraphics[width=0.31\textwidth]{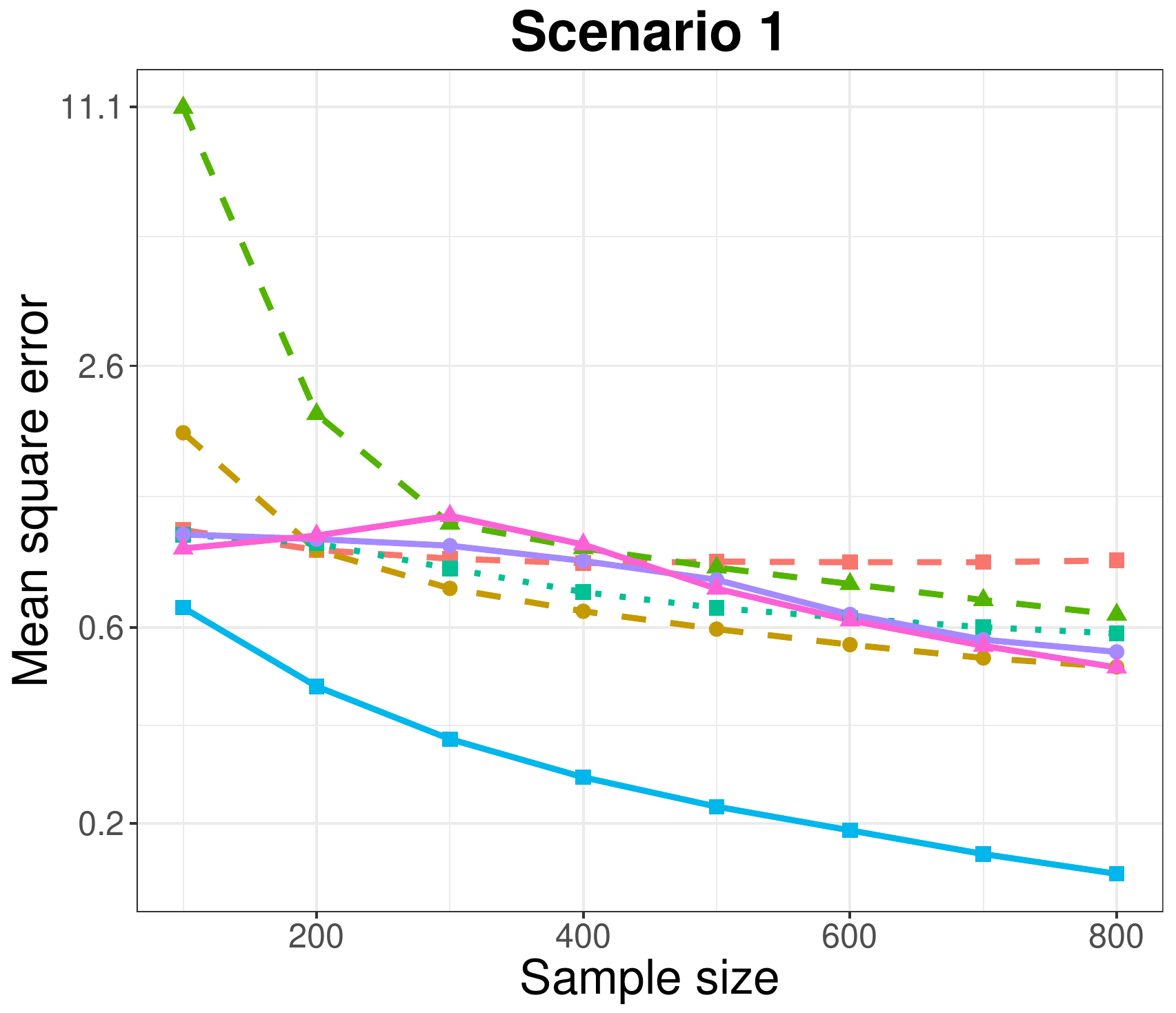}
	\includegraphics[width=0.31\textwidth]{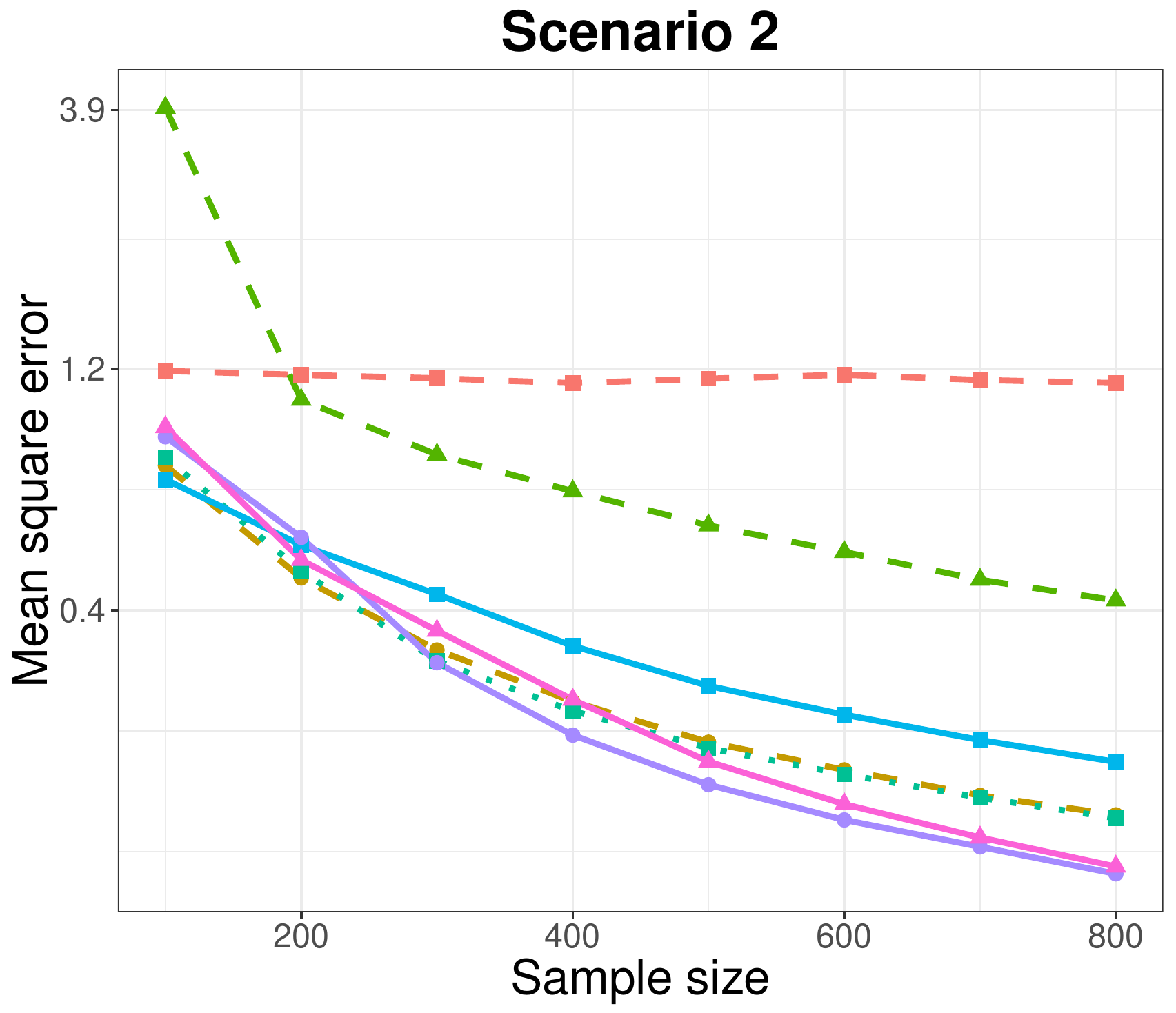}
	\includegraphics[width=0.31\textwidth]{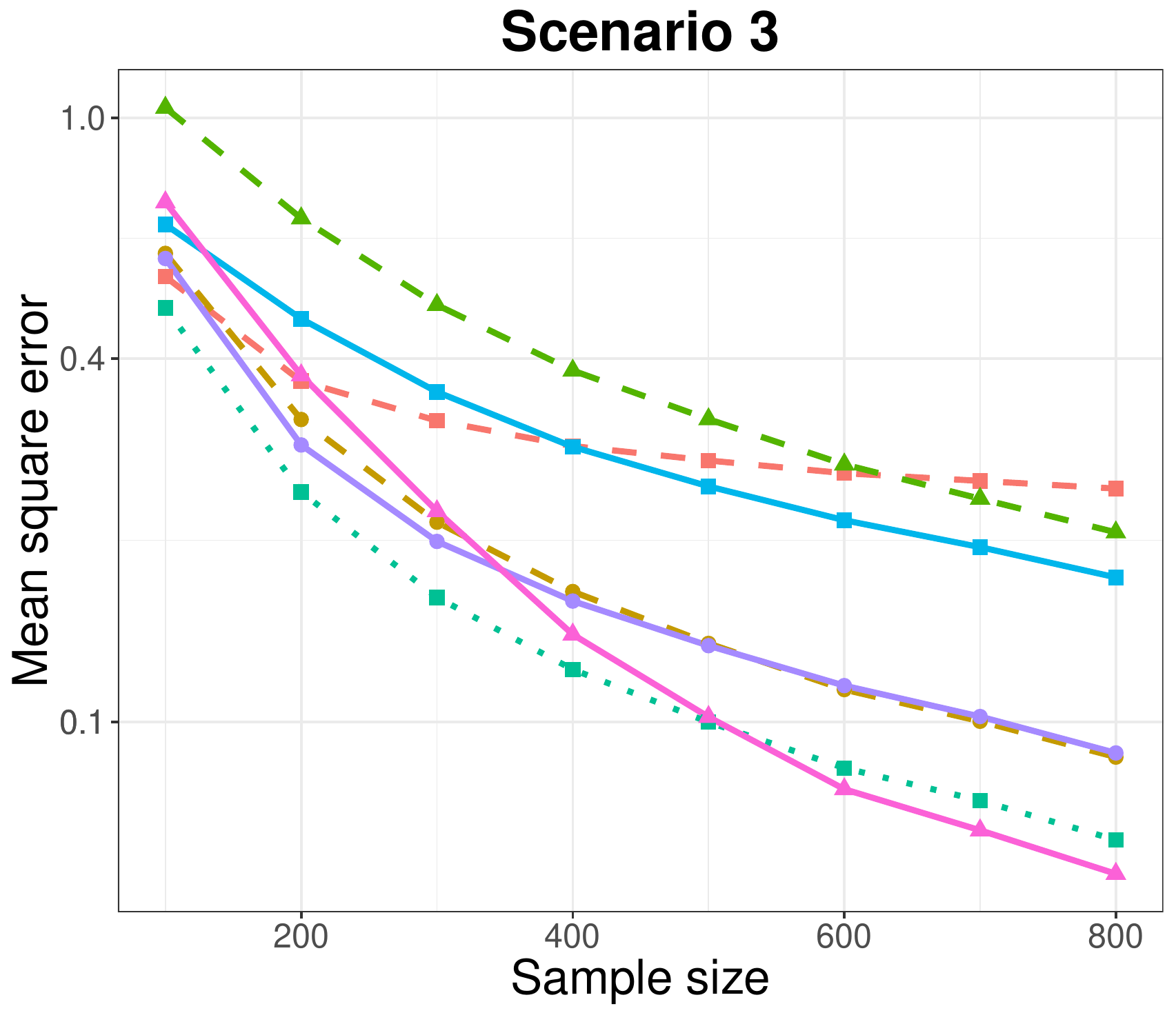}
	\includegraphics[width=0.31\textwidth]{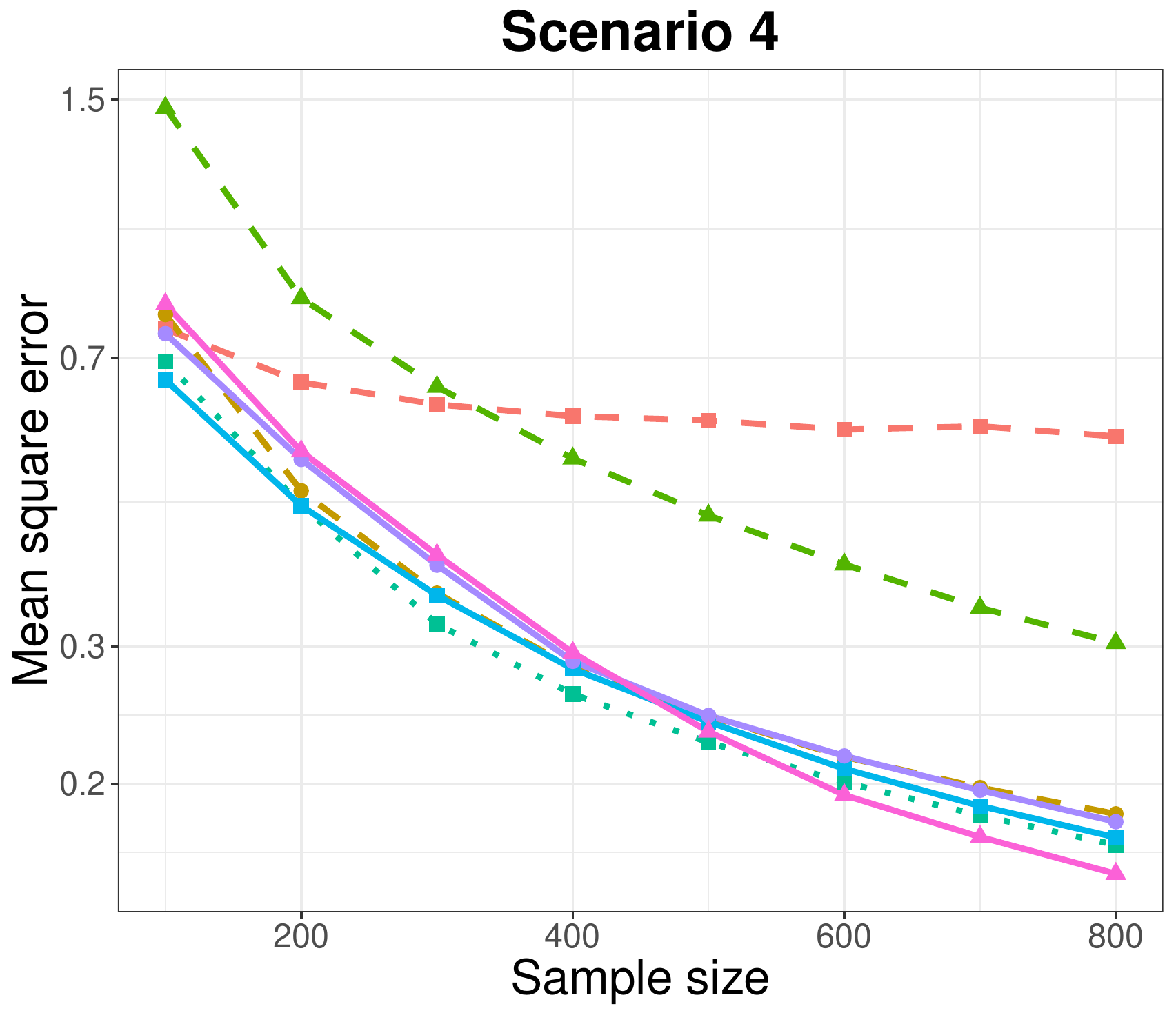}
	\includegraphics[width=0.31\textwidth]{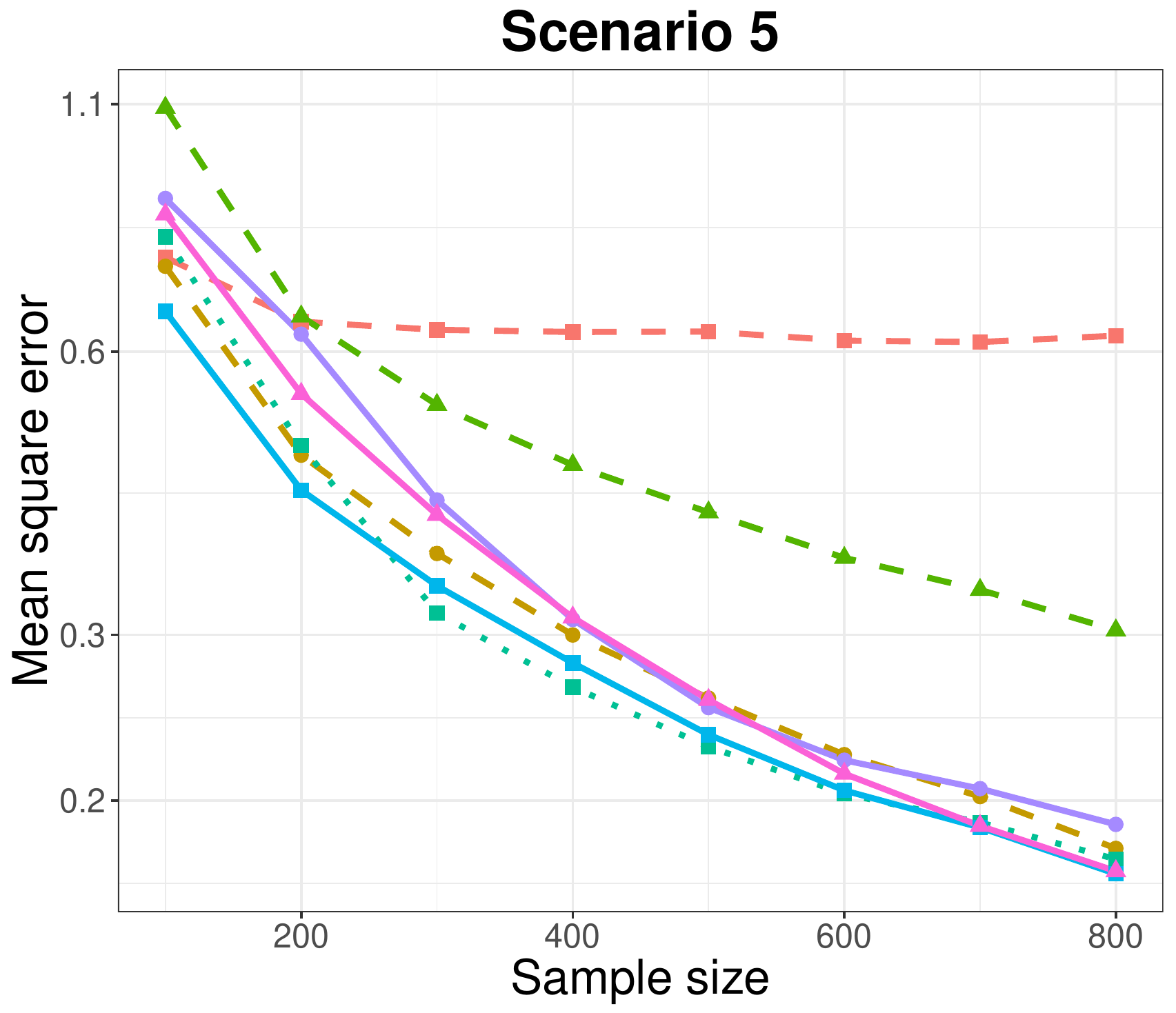}
	\caption{Plots of MSE versus sample size for each of five scenarios for $p=100$, averaged over 500 replications. The line types and colors are the same as in Figure~\ref{fig:MSEp6}.}
	\label{fig:MSEp100}
\end{figure}

In Figures~\ref{fig:MSEp6} and \ref{fig:MSEp100}, we plot the MSE as a function of $n$ for the low and high-dimensional setting, respectively. For each simulation scenario, we plot the performance of SpAM for three different choices of $M$~(low, moderate and high number of basis functions, $M$). In both low- and high-dimensional settings, we observe similar relative performances between the methods, with more variability in results for the high-dimensional setting. While there is no uniformly superior method, for all, except Scenario 1, the Sobolev smoothness penalty and trend filtering of orders 1 and 2 had comparably good performances. Unsurprisingly, trend filtering of order 0 exhibits superior performance in Scenario 1, where each component is piecewise constant. In each scenario, the bias-variance trade-off of SpAM depends on the choice of $M$: too small or large values of $M$ lead to high prediction error compared to other methods. 

In Appendix~\ref{app:AdditionalFigures}, we plot examples of fitted functions for the various methods. The dependence on $M$ for SpAM, is further illustrated in  Figure~\ref{fig:SamplePlots}, where we plot functions estimated by SpAM for high-dimensional Scenario 4 with $n=500$. We observe large bias for $M=3$~(especially for the piecewise constant and linear functions) and high variance for $M=30$. In the same figure, we also plot functions estimated by the SSP; SSP estimates exhibit a similar bias to that of SpAM with $M=10$, but with a substantially smaller variance. Figure~\ref{fig:SamplePlots2} similarly plots fitted example functions for trend filtering. Trend filtering with $k=0$ estimates the piecewise constant function well, but estimating the other $f_j$'s by piecewise constant functions incurs additional variance. Trend filtering with $k=1$ and $2$ estimates all other signal functions well.

\section{Data Analysis}
\label{sec:data}

We use the methods of Section~\ref{sec:SimulationStudy} to predict the value of owner-occupied homes in the suburbs of Boston using census data from 1970. The data consists of $n=506$ measurements and $10$ covariates, and has been studied in the additive models literature~\citep{ravikumar2009sparse,lin2006component}. As done in the data analysis by \cite{ravikumar2009sparse}, we add 10 noise covariates uniformly generated on the unit interval and 10 additional noise covariates obtained by randomly permuting the original covariates. 

We fit SSP, SpAM with $M = 2$ and $3$ basis functions, and TF with orders $k=0, 1, 2$; we also fit the lasso~\cite{tibshirani1996regression}. Approximately $75\%$ of the observations are used as training set, and the mean square prediction error on the test set is reported. The final model is selected using 5-fold cross validation using the `1 standard error rule'. Results are presented for 100 splits of the data into training and test sets.

\begin{figure}
\centering
\includegraphics[width = \textwidth, clip=TRUE, trim=0mm 7mm 0mm 0mm]{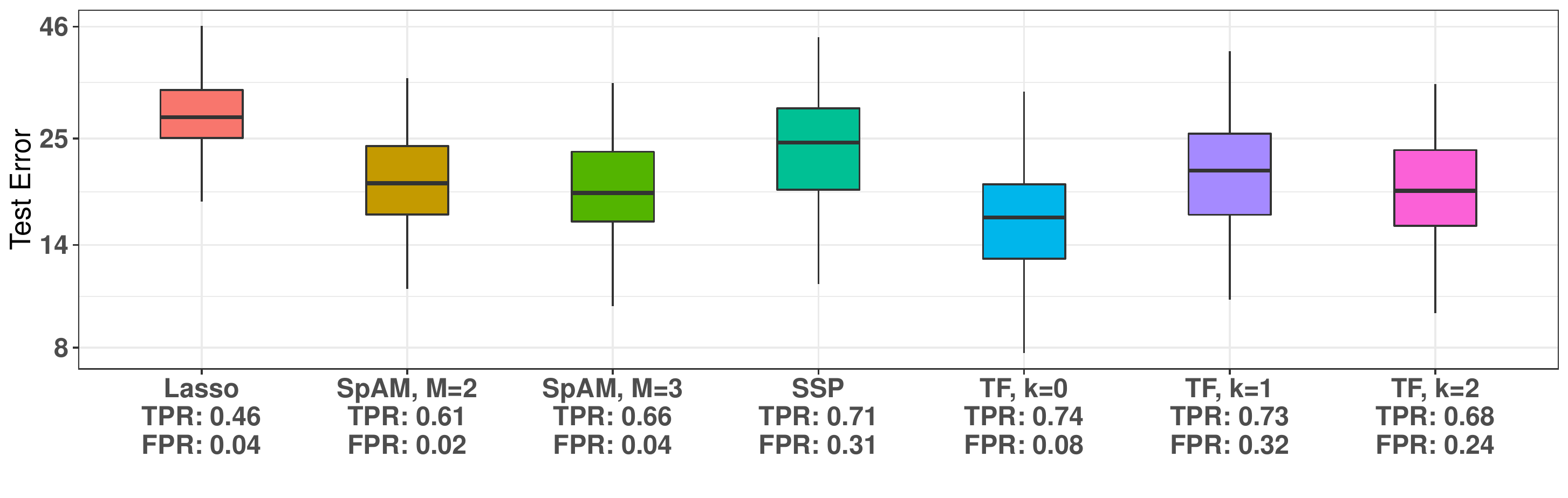}
\caption{Box-plot of test errors for 100 different train/test splits of the data for each method. The average TPR and FPR was calculated using the original 10 covariates as `signal' variables and remaining 20 as noise variables.}
\label{fig:DataAnalysis}
\end{figure}

The box-plots of test error in the test set are shown in Figure~\ref{fig:DataAnalysis}. Since we added noise variables for the purpose of this analysis, we also state the average true positive rate~(TPR) and false positive rate~(FPR) in Figure~\ref{fig:DataAnalysis}. The box-plots demonstrate superior performance of TF of order $k=0$ over other methods in terms of lowest prediction error and highest TPR. The FPR of TF with $k=0$ is also low (under 10\%). In Figure~\ref{fig:DataAnalysisExample} of Appendix~\ref{app:AdditionalFigures}, we plot fitted functions for one split of the data for lasso, SpAM with $M=3$, SSP and, TF with $k=0$ for the 10 covariates of the original dataset. A striking feature of TF fits is that many component functions are constant for extreme values of the covariates.

\section{Conclusion}
\label{sec:Conclusion}

In this paper, we introduced a general framework for non-parametric high-dimensional sparse additive models. We show that many existing proposals, such as SpAM~\citep{ravikumar2009sparse}, SPLAM~\citep{lou2016sparse}, Sobolev smoothness~\citep{meier2009high}, and trend filtering additive models~\citep{sadhanala2017additive,petersen2016fused}, fall within our framework.

We established a proximal gradient descent algorithm which has a lasso-like per-iteration complexity for certain choices of the structural penalty.  Our theoretical analyses in Section~\ref{sec:TheoreticalResults} showed both fast rates, which match minimax rates under Gaussian noise, as well as slow rates, which only require a few weak assumptions. 

The \texttt{R} package \name , available on \url{https://github.com/asadharis/GSAM}, implements the methods described in this paper.

\bibliographystyle{plainnat}
\bibliography{bibfileah/refnew}



\section*{Supplementary material}

# Supplementary Material for "Generalized Sparse Additive Models"

Asad Haris[*]     Noah Simon[†]     Ali Shojaie[‡]

March 11, 2019

## A  Block Coordinate Descent Algorithm and Additional Figures

Algorithm A.1 is the block coordinate descent algorithm that can be used to estimate `PGSAME`s.

Figure A.1 shows estimated functions for SpAM with 3,10 and 30 basis functions, and for SSP.

Figure A.2 shows estimated functions for trend filtering of orders $k = 0, 1, 2$.

Figure A.3 shows estimated functions for the various methods for the analysis of Boston housing data,

## B  Proof of Results in Section 2.2

*Proof of Lemma 1.* (a) $\Leftrightarrow$ (b). By a simple calculation we see that $P_{semi}^2$ is pathwise differentiable at **0**, and its derivative is given by $\frac{\partial}{\partial \varepsilon} P_{semi}^2(\varepsilon h)\big|_{\varepsilon=0} = \frac{\partial}{\partial \varepsilon} \varepsilon^2 P_{semi}^2(h)\big|_{\varepsilon=0} = 0$ for any function $h$. Now for any direction $h$, we can calculate a subdifferential of the objective in (4) evaluated at $f = \mathbf{0}$:

$$\frac{\partial}{\partial \varepsilon}\left[n^{-1}\sum_{i=1}^n (y_i - \varepsilon h(x_i))^2 + \lambda^2 P_{semi}^2(\varepsilon h) + \lambda \|\varepsilon h\|_n\right]_{\varepsilon=0} \ni -n^{-1}\sum_i y_i h(x_i) + \lambda U \|h\|_n \equiv \delta_{quad}(U),$$

for any $U \in [-1, 1]$. By the sub-gradient conditions, $\widehat{f}_1 = \mathbf{0}$ if and only if for every $h$, $\delta_{quad}(U) = 0$ for some $U \in [-1, 1]$. For any given $h$, we have that

$$\delta_{quad}(U) = 0 \Leftrightarrow -n^{-1}\sum_i y_i h(x_i) + \lambda U \|h\|_n = 0 \Leftrightarrow \left|n^{-1}\sum_i y_i \frac{h(x_i)}{\|h\|_n}\right| = \lambda |U|.$$

From this we see that for a given $h$, $\delta_{quad}(U) = 0$ for some $U \in [-1, 1]$ if and only if $\left|n^{-1}\sum_i y_i h(x_i)/\|h\|_n\right| \leq \lambda$. It follows that, $\widehat{f}_1 = \mathbf{0}$ if and only if $\left|n^{-1}\sum_i y_i h(x_i)/\|h\|_n\right| \leq \lambda$ for every direction $h$.

---

[*]asad.haris@mail.mcgill.ca, Department of Epidemiology, Biostatistics and Occupational Health, McGill University

[†]nrsimon@uw.edu, Department of Biostatistics, University of Washington

[‡]ashojaie@uw.edu, Department of Biostatistics, University of Washington

1

**Algorithm A.1** Block Coordinate Descent for Least Squares Loss

1: Initialize $f_1^0, \ldots f_p^0 \leftarrow \mathbf{0}, \beta^0 \leftarrow 0, \boldsymbol{r} \leftarrow \boldsymbol{y}, k \leftarrow 1$
2: **while** $k \leq max\_iter$ **and** not converged **do**
3:     Update
$$\beta^k \leftarrow n^{-1} \sum_{i=1}^n r_i, \quad \boldsymbol{r} \leftarrow \boldsymbol{r} - \beta^k \mathbf{1}.$$
4:     **for** $j = 1, \ldots, p$ **do**
5:         Set $\boldsymbol{r}_{-j}$ as
$$r_{-j,i} = r_i + f_j^{k-1}(x_{ij}).$$
6:         Update
$$f_j^k \leftarrow \left(1 - t\lambda/\|f_j^{inter}\|_n\right)_+ f_j^{inter},$$
        where
$$f_j^{inter} \leftarrow \operatorname*{argmin}_{f \in \mathcal{F}} \frac{1}{2}\left\|\boldsymbol{r}_{-j} - f\right\|_n^2 + t\lambda^2 P_{st}(f). \tag{A.1}$$
7:         Update $\boldsymbol{r}$ to
$$r_i \leftarrow r_{-j,i} + f_j^k(x_{ij}).$$
8:     **end for**
9: **end while**
10: **return** $\beta^k, f_1^k, \ldots, f_p^k$

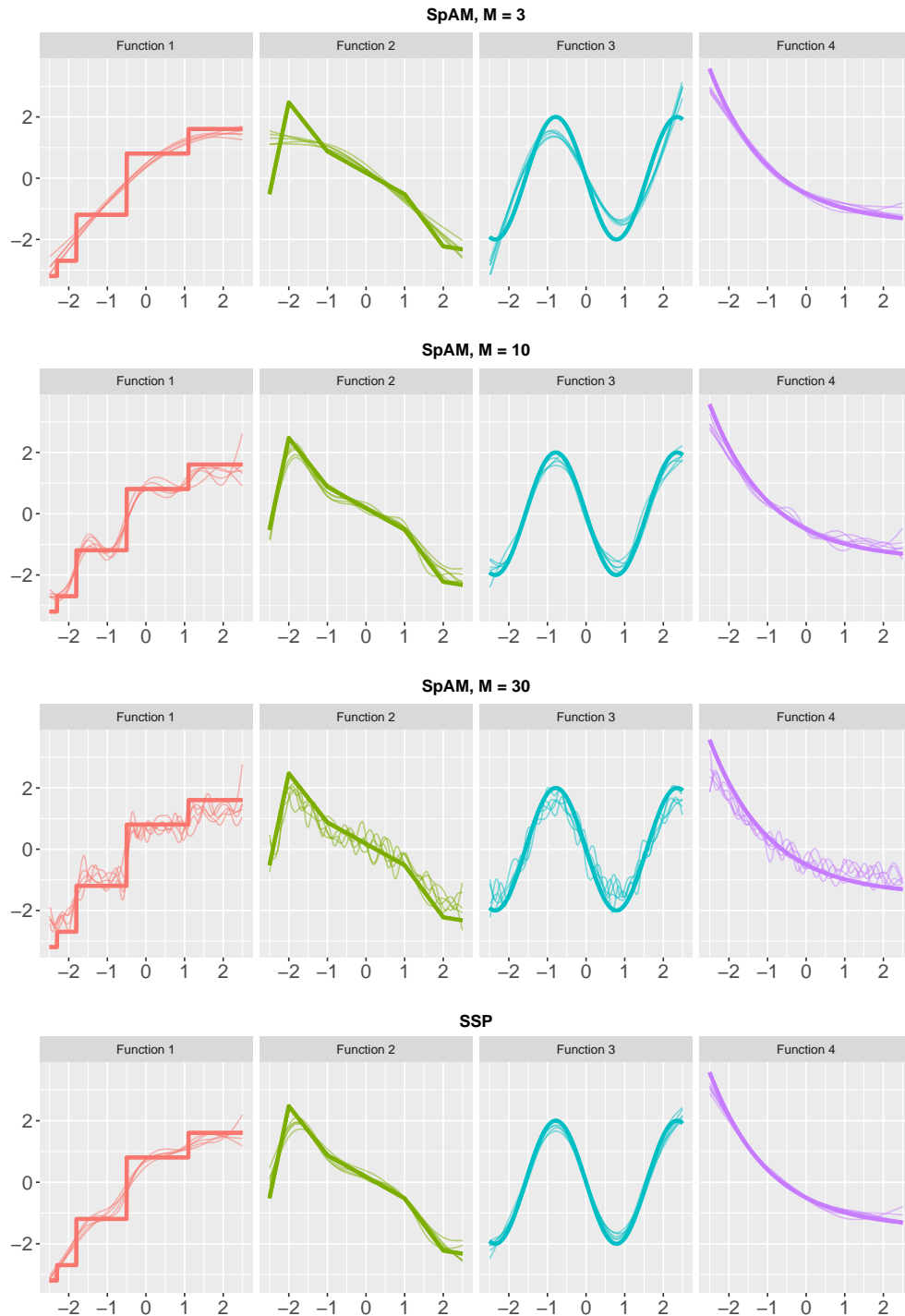

Figure A.1: Examples of estimated signal functions by SpAM [Ravikumar et al., 2009] and Sobolev Smoothness Penalty [Meier et al., 2009] for Scenario 4.

3

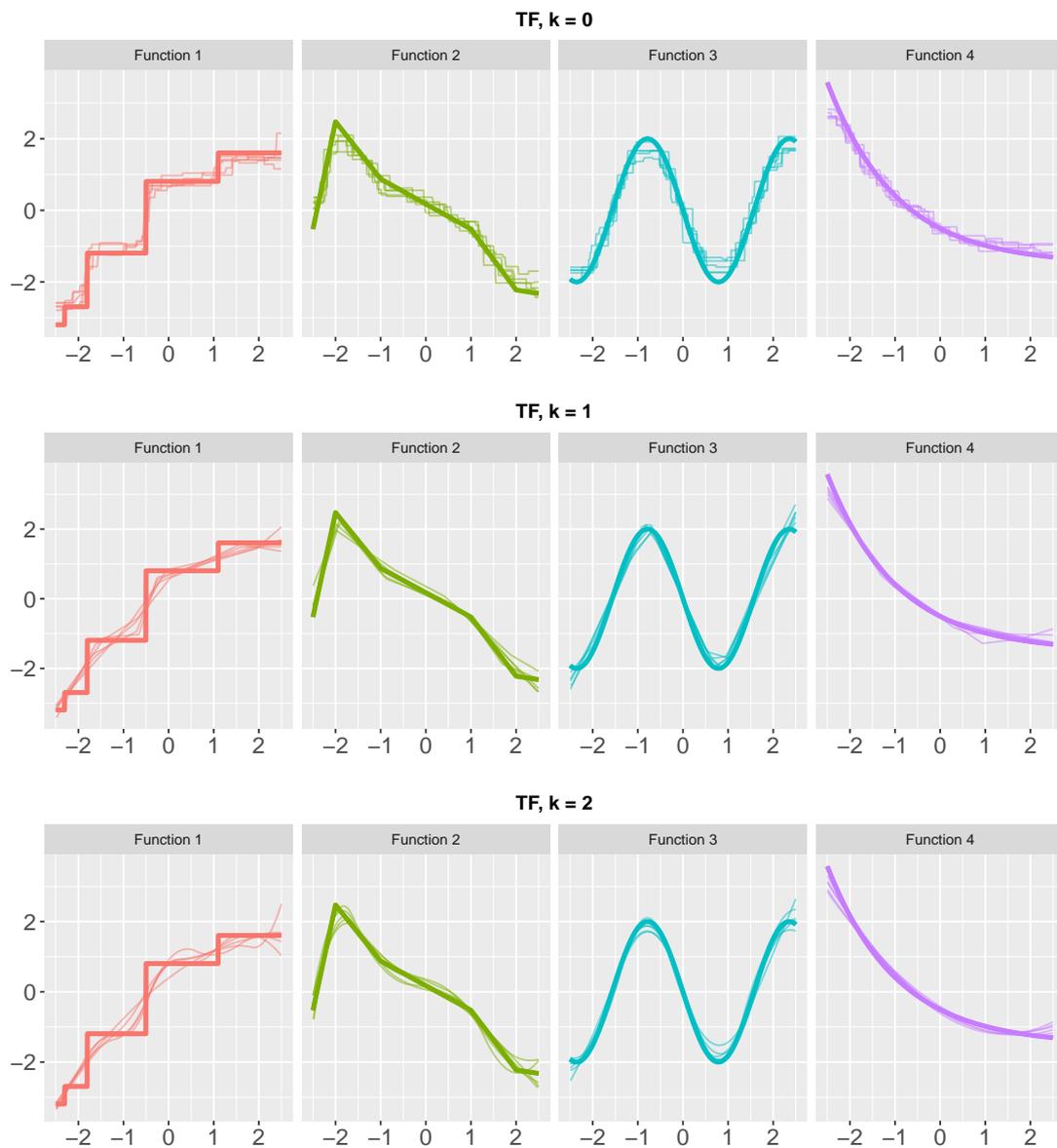

Figure A.2: Examples of estimated signal functions by Trend Filtering [Sadhanala and Tibshirani, 2018] for Scenario 4.

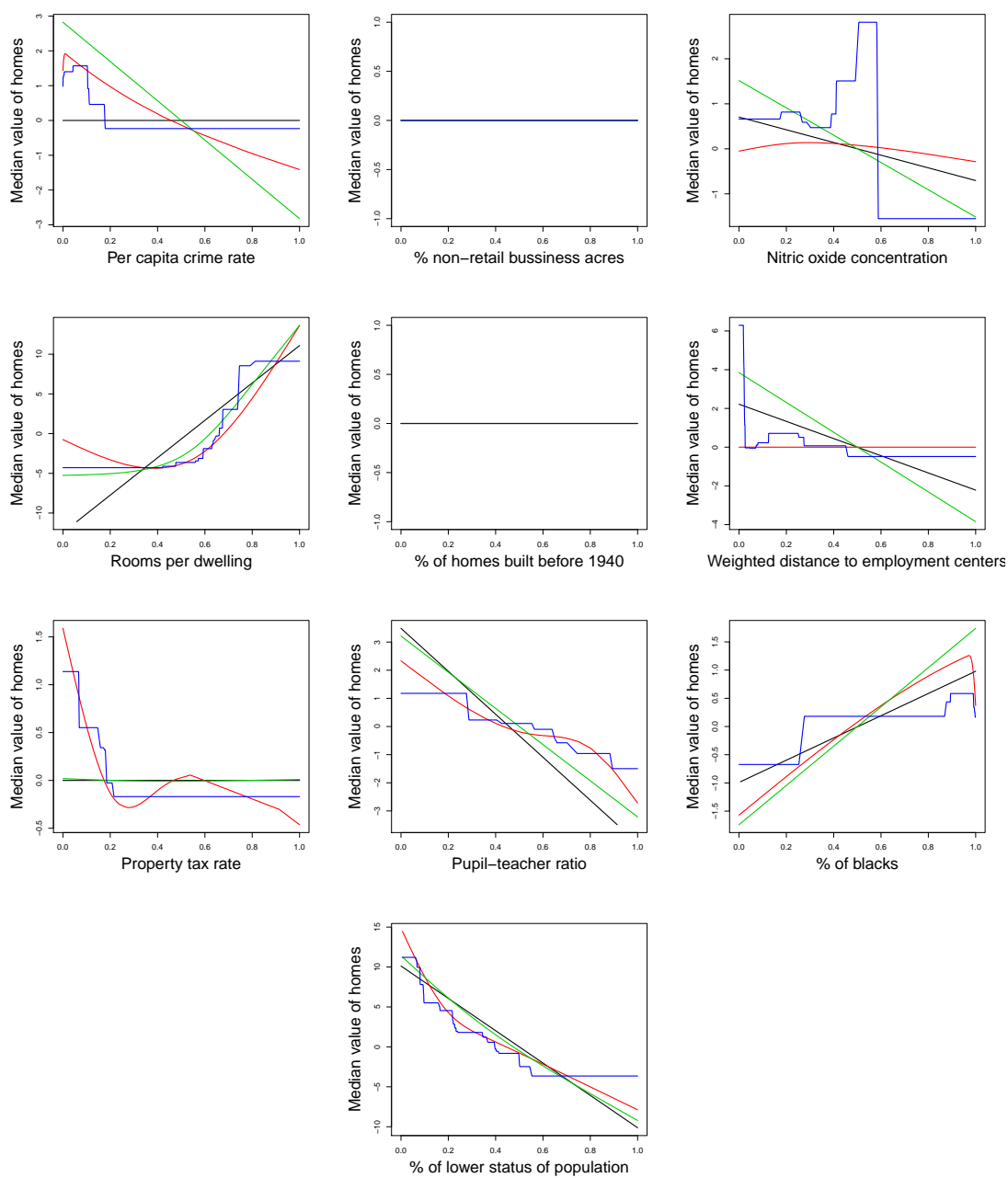

Figure A.3: Plots of fitted functions for the original 10 covariates for a single split of the data into training and test sets for lasso (———), SpAM (———) with $M = 3$ basis functions, SSP (———) and, TF (———) of order $k = 0$.

(b) ⇒ (c). Since $\left|n^{-1}\sum_i y_i h(x_i)/\|h\|_n\right| \le \lambda$ holds for every direction $h$, we simply consider the special direction such that $h(x_i) = y_i$. This implies (c).

(b) ⇐ (c). Assuming that $\|\boldsymbol{y}\|_n \le \lambda$ then, for every $h$

$$\left|n^{-1}\sum_i y_i \frac{h(x_i)}{\|h\|_n}\right| \le \|\boldsymbol{y}\|_n \frac{\|h\|_n}{\|h\|_n} \le \|\boldsymbol{y}\|_n \le \lambda.$$

□

*Proof of Lemma 2.* (a) ⇔ (b). By the absolute homogeneity of semi-norms,

$$\frac{\partial}{\partial \varepsilon} P_{semi}(\varepsilon h)\Big|_{\varepsilon=0} = \frac{\partial}{\partial \varepsilon}|\varepsilon| P_{semi}(h)\Big|_{\varepsilon=0}.$$

Hence the subdifferential evaluated at $f = \boldsymbol{0}$, in the direction of $h$, is

$$\frac{\partial}{\partial \varepsilon}\left[n^{-1}\sum_{i=1}^n (y_i - \varepsilon h(x_i))^2 + \lambda^2 P_{semi}(\varepsilon h) + \lambda \|\varepsilon h\|_n\right]_{\varepsilon=0} \ni -n^{-1}\sum_i y_i h(x_i) + \lambda^2 V P_{semi}(h) + \lambda U \|h\|_n$$

$$\equiv \delta_{semi}(U, V),$$

for any $(U, V) \in [-1, 1]^2$. As in the previous lemma, $\widehat{f}_2 = \boldsymbol{0}$ if and only if for every $h$, there exists $(U, V) \in [-1, 1]^2$ such that $\delta_{semi}(U, V) = 0$. For a given $h$, we have that

$$\delta_{semi}(U, V) = 0 \Leftrightarrow n^{-1}\sum_i y_i h(x_i) - \lambda^2 V P_{semi}(h) = \lambda U \|h\|_n$$

$$\Leftrightarrow \left|n^{-1}\sum_i y_i \frac{h(x_i)}{\|h\|_n} - \lambda^2 V \frac{P_{semi}(h)}{\|h\|_n}\right| = \lambda |U|.$$

Thus for a given $h$, $\delta_{semi}(U, V) = 0$ for some $(U, V) \in [-1, 1]^2$ if and only if

$$\left|n^{-1}\sum_i y_i \frac{h(x_i)}{\|h\|_n} - \lambda^2 V \frac{P_{semi}(h)}{\|h\|_n}\right| \le \lambda,$$

for some $V \in [-1, 1]$. This proves the first part.

Now we will show that $\|\boldsymbol{y}\|_n \le \lambda$ implies (b). That is, for every $h$, there exists some $V \in [-1, 1]$ such that

$$\left|n^{-1}\sum_i y_i \frac{h(x_i)}{\|h\|_n} - \lambda^2 V \frac{P_{semi}(h)}{\|h\|_n}\right| \le \lambda.$$

We can simply take $V = 0$ for every $h$, which reduces the above inequality to the one seen in the proof of Lemma 1. Thus $\|\boldsymbol{y}\|_n \le \lambda \Rightarrow \widehat{f}_2 = \boldsymbol{0}$. □

## C  Proof of Results in Section 3

*Proof of Lemma 3.* If $\widetilde{f} = 0$, then $\widehat{f} = 0$ is trivially the solution to (12). Thus, throughout this proof, we consider $\widetilde{f} \ne 0$.

**Case 1:** $\|\widetilde{f}\|_n \ge \lambda_2$. In this case $c\widehat{f} = \widetilde{f}$ where $c = (1 - \lambda_2/\|\widetilde{f}\|_n)^{-1}$. Let $f_T \in \mathcal{F}$ be some arbitrary function and define the function $h = f_T - \widehat{f}$. We will show that along the path $\widehat{f} + \varepsilon h$ for all $\varepsilon \in [0, 1]$, the objective

$$\frac{1}{2}\left\|r - (\widehat{f} + \varepsilon h)\right\|_n^2 + \lambda_1 P_{st}\left(\widehat{f} + \varepsilon h\right) + \lambda_2\|\widehat{f} + \varepsilon h\|_n \tag{C.1}$$

is minimized at $\varepsilon = 0$. We begin by noting that
$$\frac{1}{2}\left\|r - (\widetilde{f} + \varepsilon ch)\right\|_n^2 + \lambda_1 P_{st}\left(\widetilde{f} + \varepsilon ch\right),$$
is minimized at $\varepsilon = 0$ because
$$\widetilde{f} + \varepsilon ch = \widetilde{f} + \varepsilon c f_T - \varepsilon c \widehat{f} = (1 - \varepsilon)\widetilde{f} + \varepsilon c f_T \in \mathcal{F},$$
for all $\varepsilon \in [0, 1]$ since $\mathcal{F}$ is a convex cone. By the sub-gradient condition, we have
$$-\langle r - \widetilde{f}, ch \rangle_n + \lambda_1 \vartheta_1 = 0,$$
for some $\vartheta_1 \in \partial P_{st}\left(\widetilde{f} + \varepsilon ch\right)\Big|_{\varepsilon=0}$, or equivalently
$$c\left[-\langle r - c\widehat{f}, h \rangle_n + \lambda_1 \vartheta_2\right] = 0,$$
for some $\vartheta_2 \in \partial P_{st}\left(\widehat{f} + \varepsilon h\right)\Big|_{\varepsilon=0}$.

At $\widehat{f} + \varepsilon h$, one possible sub-gradient of the objective (C.1) is
$$-\langle r - \widehat{f}, h \rangle_n + \lambda_1 \vartheta_2 + \lambda_2 \frac{\langle \widehat{f}, h \rangle_n}{\|\widehat{f}\|_n}.$$
By the definition of $c$, we have that $\lambda_2/\|\widehat{f}\|_n = c\lambda_2/\|\widetilde{f}\|_n = c(1 - 1/c) = c - 1$, and thus the above sub-gradient is
$$-\langle r - \widehat{f}, h \rangle_n + \lambda_1 \vartheta_2 + (c-1)\langle \widehat{f}, h \rangle_n = -\langle r - c\widehat{f}, h \rangle_n + \lambda_1 \vartheta_2 = 0.$$
Thus we have shown that the objective function (C.1) is minimized at $\varepsilon = 0$. Since $f_T$ was an arbitrary function, we conclude that $\widehat{f}$ is the solution of (12).

**Case 2:** $\|\widetilde{f}\|_n < \lambda_2$. In this case we will show that $\widehat{f} \equiv 0$. For this, we consider the path $\varepsilon f_T$ for $\varepsilon \in [0, 1]$ for an arbitrary $f_T \in \mathcal{F}$. We will show that the function
$$\frac{1}{2}\|r - \varepsilon f_T\|_n^2 + \lambda_1 P_{st}\left(\varepsilon f_T\right) + \lambda_2 \|\varepsilon f_T\|_n, \tag{C.2}$$
is minimized at $\varepsilon = 0$ and since $f_T$ is arbitrary that will complete the proof.

As in the previous case, we begin by looking at the sub-gradient conditions for $\widetilde{f}$. The expression
$$\frac{1}{2}\left\|r - (\widetilde{f} + \varepsilon f_T)\right\|_n^2 + \lambda_1 P_{st}\left(\widetilde{f} + \varepsilon f_T\right),$$
is minimized at $\varepsilon = 0$ by definition of $\widetilde{f}$. This leads us to the sub-gradient condition
$$-\langle r - \widetilde{f}, f_T \rangle_n + \lambda_1 \vartheta_1 = 0 \Leftrightarrow \vartheta_1 = \frac{\langle r - \widetilde{f}, f_T \rangle_n}{\lambda_1}.$$
Now we describe the the sub-gradient conditions for (C.2). All sub-gradients of (C.2) at $\varepsilon = 0$ are given by
$$-\langle r, f_T \rangle_n + \nu_1 \lambda_1 P_{st}(f_T) + \nu_2 \lambda_2 \|f_T\|_n, \tag{C.3}$$

7

for real values $(\nu_1, \nu_2) \in [-1,1]^2$. To complete the proof we need to find $\nu_1$ and $\nu_2$ such that (C.3) is 0 and, $(\nu_1, \nu_2) \in [-1,1]^2$. Setting $\nu_1 = \vartheta_1/P_{st}(f_T)$ and $\nu_2 = \langle \widetilde{f}, f_T \rangle/(\lambda_2 \|f_T\|_n)$ clearly makes (C.3) 0 and so we need only prove that our choice of $\nu_1$ and $\nu_2$ lie within the interval $[-1,1]$.

Showing $|\nu_1| \leq 1$ is equivalent to $|\vartheta_1| \leq P_{st}(f_T)$. $\vartheta_1$ is a member of the sub-gradient set

$$\partial P_{st}\left(\widetilde{f} + \varepsilon f_T\right)\Big|_{\varepsilon=0} = \left\{u \geq 0 : P_{st}(\widetilde{f} + \eta f_T) - P(\widetilde{f}) \geq u \times \eta \quad \forall \eta \geq 0\right\}.$$

Thus $\vartheta_1$ must satisfy the inequality

$$\vartheta_1 \eta \leq P_{st}(\widetilde{f} + \eta f_T) - P_{st}(\widetilde{f})$$
$$\leq P_{st}(\widetilde{f}) + \eta P_{st}(f_T) - P_{st}(\widetilde{f}) = \eta P_{st}(f_T),$$

where the second inequality holds because $P_{st}$ is a semi-norm. This proves that $|\vartheta_1| \leq P_{st}(f_T)$.

Showing $|\nu_2| \leq 1$ is easier and follows by definition:

$$|\nu_2| = \frac{|\langle \widetilde{f}, f_T \rangle|}{\lambda_2 \|f_T\|_n} \leq \frac{\|\widetilde{f}\|_n \|f_T\|_n}{\lambda_2 \|f_T\|_n} = \frac{\|\widetilde{f}\|_n}{\lambda_2},$$

which is less than 1 since $\|\widetilde{f}\|_n < \lambda_2$.

$\square$

*Prof of Lemma 4.* Consider an arbitrary direction $\widehat{f}_\lambda + \varepsilon h$ for some function $h$ and $\varepsilon$ in an open interval. We first consider the case $P_{st}(\widehat{f}_\lambda) \neq 0$. In this case if the directional derivative $\nabla_h P_{st}^\nu(\widehat{f})$ exists then so does the directional derivative of $P_{st}(\widehat{f})$ and is given by

$$\nabla_h P_{st}(\widehat{f}_\lambda) = \frac{\nabla_h P_{st}^\nu(\widehat{f}_\lambda)}{\nu P_{st}^{\nu-1}(\widehat{f}_\lambda)}.$$

This follows from standard arguments for derivative of power functions. Here we present the simple case of integer valued $\nu > 1$.

$$\nabla_h P_{st}(\widehat{f}_\lambda) = \lim_{\varepsilon \to 0} \frac{P_{st}(\widehat{f}_\lambda + \varepsilon h) - P_{st}(\widehat{f}_\lambda)}{\varepsilon}$$
$$= \lim_{\varepsilon \to 0} \frac{P_{st}(\widehat{f}_\lambda + \varepsilon h) - P_{st}(\widehat{f}_\lambda)}{\varepsilon} \times \frac{\sum_{l=1}^\nu \{P_{st}(\widehat{f}_\lambda + \varepsilon h)\}^{\nu-l}\{P_{st}(\widehat{f}_\lambda)\}^{l-1}}{\sum_{l=1}^\nu \{P_{st}(\widehat{f}_\lambda + \varepsilon h)\}^{\nu-l}\{P_{st}(\widehat{f}_\lambda)\}^{l-1}}$$
$$= \lim_{\varepsilon \to 0} \frac{P_{st}^\nu(\widehat{f}_\lambda + \varepsilon h) - P_{st}^\nu(\widehat{f}_\lambda)}{\varepsilon} \times \lim_{\varepsilon \to 0} \frac{1}{\sum_{l=1}^\nu \{P_{st}(\widehat{f}_\lambda + \varepsilon h)\}^{\nu-l}\{P_{st}(\widehat{f}_\lambda)\}^{l-1}}$$
$$= \nabla_h P_{st}^\nu(\widehat{f}_\lambda) \times \frac{1}{\sum_{l=1}^\nu \{P_{st}(\widehat{f}_\lambda)\}^{\nu-1}} = \frac{\nabla_h P_{st}^\nu(\widehat{f}_\lambda)}{\nu P_{st}^{\nu-1}(\widehat{f}_\lambda)}.$$

Now, by the gradient condition, for $f_\varepsilon = \widehat{f}_\lambda + \varepsilon h$,

$$\frac{\partial}{\partial \epsilon}\left[\frac{1}{2}\|r - f_\varepsilon\|_n^2 + \lambda P_{st}(f_\varepsilon)\right]_{\epsilon=0} = -\langle r - \widehat{f}_\lambda, h\rangle_n + \lambda \frac{\nabla_h P_{st}^\nu(\widehat{f}_\lambda)}{\nu P_{st}^{\nu-1}(\widehat{f}_\lambda)} = 0, \quad (C.4)$$

Similarly, for the path $f_{\widetilde{\varepsilon}} = \widetilde{f}_{\widetilde{\lambda}} + \widetilde{\varepsilon} h$, by the gradient condition,

$$\frac{\partial}{\partial \widetilde{\varepsilon}}\left[\frac{1}{2}\|r - f_{\widetilde{\varepsilon}}\|_n^2 + \widetilde{\lambda} P_{st}^\nu(f_{\widetilde{\varepsilon}})\right]_{\widetilde{\varepsilon}=0} = -\langle r - \widetilde{f}_{\widetilde{\lambda}}, h\rangle_n + \widetilde{\lambda} \nabla_h P_{st}^\nu\left(\widetilde{f}_{\widetilde{\lambda}}\right) = 0.$$

8

This is exactly the optimality condition (C.4) with $\widetilde{\lambda} = \lambda(\nu P_{st}^{\nu-1}(\widehat{f}_\lambda))^{-1}$. Thus, if $\nu\widetilde{\lambda}P_{st}^{\nu-1}(\widetilde{f}_{\widetilde{\lambda}}) = \lambda$ then $\widehat{f}_\lambda = \widetilde{f}_{\widetilde{\lambda}}$.

Now to show the case $P_{st}(\widehat{f}) = 0$, we need to find conditions for which the objective

$$\frac{1}{2}\|f_{interp} - f\|_n^2 + \lambda P_{st}(f) \tag{C.5}$$

is minimized by $\widehat{f} = f_{null}$. We consider the functions $f_{h,\varepsilon} = (1-\varepsilon)f_{null} + \varepsilon h$ for $\varepsilon \in [0,1]$ and show that for all $h \in \mathcal{F}$, the objective

$$\frac{1}{2}\|f_{interp} - f_{h,\varepsilon}\|_n^2 + \lambda P_{st}(f_{h,\varepsilon}) \tag{C.6}$$

is minimized at $\varepsilon = 0$, if and only if $\lambda \geq P_{st}^*(f_{interp} - f_{null})$.

To see this, note that all subgradients of (C.6) at $\varepsilon = 0$, are of the form

$$\langle f_{interp} - f_{null}, f_{null} - h\rangle_n + \lambda \kappa P_{st}(h),$$

for $\kappa \in [-1, 1]$. For 0 to be a sub-gradient of (C.6) we need to have

$$\lambda \kappa P_{st}(h) = \langle f_{interp} - f_{null}, h - f_{null}\rangle_n. \tag{C.7}$$

Consequently, (C.6) is minimized at $\varepsilon = 0$ if and only if for all $h \in \mathcal{F}$

$$\langle f_{interp} - f_{null}, h - f_{null}\rangle_n \leq \lambda P_{st}(h). \tag{C.8}$$

Using the decomposition $h = h_0 + h_\perp \in \mathcal{F}_1 \oplus \mathcal{F}_2$, the above condition becomes

$$\frac{\langle f_{interp} - f_{null}, h_0 - f_{null}\rangle_n}{P_{st}(h_\perp)} + \frac{\langle f_{interp} - f_{null}, h_\perp\rangle_n}{P_{st}(h_\perp)} \leq \lambda. \tag{C.9}$$

Now if $f_{interp} - f_{null} \in \mathcal{F}_2$, then the first part of the LHS is 0 and the second part is bounded above by $P_{st}^*(f_{interp} - f_{null})$. To complete the proof we show that $f_{interp} - f_{null}$ is, infact, a member of $\mathcal{F}_2$. For this, it suffices to show that $\langle f_{interp} - f_{null}, f_{null} - h_{null}\rangle_n = 0$ for all $h_{null} \in \mathcal{F}_1$. We know that $f_{null}$ is the solution to the problem

$$\underset{f \in \mathcal{F}_1}{\text{minimize}} \frac{1}{2}\|f_{interp} - f\|_n^2;$$

in other words, for all $h_{null} \in \mathcal{F}_1$, the expression

$$\frac{1}{2}\|f_{interp} - (1-\varepsilon)f_{null} - \varepsilon h_{null}\|_n^2, \tag{C.10}$$

is minimized by $\varepsilon = 0$. Equivalently (by the gradient condition), for all $h_{null} \in \mathcal{F}_1$,

$$\langle f_{interp} - f_{null}, f_{null} - h_{null}\rangle_n = 0. \tag{C.11}$$

□

9

# D   Proofs of Results in Section 4.2

In this section we present the proof Theorem 1 and 3 for the sake of completeness. The arguments presented here are only a slight modification to those of Bühlmann and van de Geer [2011] for proving LASSO rates. One notable difference is that we explicitly handle an unpenalized intercept term; another is our handling of the structural penalty, $P_{st}(\cdot)$. Throughout the proofs, we will utilize the so-called *basic inequalities*. Hence, for the sake of convenience, we state and prove these basic inequalities as a separate lemma.

**Lemma D.1** (Basic Inequality). *Let $\widehat{f}(\boldsymbol{x}) = \widehat{\beta} + \sum_{j=1}^{p} \widehat{f}_j(x_j)$ be as defined in (19), and let $f^*(\boldsymbol{x}) = \beta^* + \sum_{j=1}^{p} f_j^*(x_j)$ be an arbitrary additive function with $\beta^* \in \mathcal{R}$ and $f_j^* \in \mathcal{F}$. Then we have the following basic inequality*

$$\mathcal{E}(\widehat{f}) + \lambda I(\widehat{f}) \leq -\left[\nu_n(\widehat{f}) - \nu_n(f^*)\right] + \lambda I(f^*) + \mathcal{E}(f^*). \tag{D.1}$$

*If we further assume that $-\ell(\cdot)$ and $P_{st}(\cdot)$ are convex, then for all $t \in (0,1)$ and $\widetilde{f} = t\widehat{f} + (1-t)f^*$ we have the following basic inequality*

$$\mathcal{E}(\widetilde{f}) + \lambda I(\widetilde{f}) \leq -\left[\nu_n(\widetilde{f}) - \nu_n(f^*)\right] + \lambda I(f^*) + \mathcal{E}(f^*). \tag{D.2}$$

*Proof.* For the first inequality, note that

$$-\mathbb{P}_n \ell\left(\widehat{f}\right) + \lambda I(\widehat{f}) \leq -\mathbb{P}_n \ell(f^*) + \lambda I(f^*),$$

which is equivalent to

$$\lambda I(\widehat{f}) \leq \mathbb{P}_n \ell\left(\widehat{f}\right) - \mathbb{P}_n \ell(f^*) + \lambda I(f^*)$$
$$\Leftrightarrow \mathbb{P}(\ell(f^*)) - \mathbb{P}(\ell(\widehat{f})) + \lambda I(\widehat{f}) \leq \mathbb{P}_n \ell\left(\widehat{f}\right) - \mathbb{P}(\ell(\widehat{f})) - \mathbb{P}_n \ell(f^*) + \mathbb{P}(\ell(f^*)) + \lambda I(f^*)$$
$$\Leftrightarrow \mathbb{P}(\ell(f^*)) - \mathbb{P}(\ell(\widehat{f})) + \lambda I(\widehat{f}) \leq -\left[(\mathbb{P}_n - \mathbb{P})(-\ell(\widehat{f})) - (\mathbb{P}_n - \mathbb{P})(-\ell(f^*))\right] + \lambda I(f^*)$$
$$\Leftrightarrow \mathbb{P}(\ell(f^*)) - \mathbb{P}(\ell(\widehat{f})) + \lambda I(\widehat{f}) \leq -\left[\nu_n(\widehat{f}) - \nu_n(f^*)\right] + \lambda I(f^*)$$
$$\Leftrightarrow \mathbb{P}(\ell(f^*)) - \mathbb{P}(\ell(f^0)) + \mathbb{P}(\ell(f^0)) - \mathbb{P}(\ell(\widehat{f})) + \lambda I(\widehat{f}) \leq -\left[\nu_n(\widehat{f}) - \nu_n(f^*)\right] + \lambda I(f^*)$$
$$\Leftrightarrow -\mathcal{E}(f^*) + \mathcal{E}(\widehat{f}) + \lambda I(\widehat{f}) \leq -\left[\nu_n(\widehat{f}) - \nu_n(f^*)\right] + \lambda I(f^*).$$

For the second inequality we have by convexity

$$-\mathbb{P}_n \ell\left(\widetilde{f}\right) + \lambda I(\widetilde{f}) \leq t\left[-\mathbb{P}_n \ell\left(\widehat{f}\right) + \lambda I(\widehat{f})\right] + (1-t)\left[-\mathbb{P}_n \ell(f^*) + \lambda I(f^*)\right]$$
$$\leq -\mathbb{P}_n \ell(f^*) + \lambda I(f^*),$$

after which we simply need to repeat the arguments for the previous basic inequality with $\widehat{f}$ replaced by $\widetilde{f}$. □

***Proof of Theorem 1.*** Define
$$t = \frac{M^*}{M^* + I(\widehat{f} - f^*)}, \tag{D.3}$$

and $\widetilde{f} = t\widehat{f} + (1-t)f^*$. Then (D.3) implies $I(\widetilde{f} - f^*) \le M^*$ and by Lemma D.1, we obtain
$$\mathcal{E}(\widetilde{f}) + \lambda I(\widetilde{f}) \le Z_{M^*} + \lambda I(f^*) + \mathcal{E}(f^*).$$

By applying the triangle inequality $I(\widetilde{f} - f^* + f^*) \ge I(\widetilde{f} - f^*) - I(f^*)$, to the left hand side we obtain on the set $\mathcal{T}$ (where $Z_{M^*} \le \rho M^* + 2R\rho$)
$$\mathcal{E}(\widetilde{f}) + \lambda I(\widetilde{f} - f^*) \le \rho M^* + 2R\rho + 2\lambda I(f^*) + \mathcal{E}(f^*).$$

Recall the definition of $M^*$, given by
$$\rho M^* = \mathcal{E}(f^*) + 2\lambda I(f^*) + 2R\rho, \tag{D.4}$$

from which we obtain
$$\mathcal{E}(\widetilde{f}) + \lambda I(\widetilde{f} - f^*) \le 2\rho M^* \le 2\frac{\lambda}{4} M^* \Rightarrow I(\widetilde{f} - f^*) \le \frac{M^*}{2}.$$

Now by the definition of $\widetilde{f}$ we have
$$I(\widehat{f} - f^*) = \frac{I(\widetilde{f} - f^*)}{t} = I(\widetilde{f} - f^*)\left[1 + \frac{I(\widehat{f} - f^*)}{M^*}\right] \le \frac{M^*}{2} + \frac{I(\widehat{f} - f^*)}{2},$$

which implies that $I(\widehat{f} - f^*) \le M^*$. Now we can repeat the above arguments with $\widetilde{f}$ replaced by $\widehat{f}$ which gives us
$$\mathcal{E}(\widehat{f}) + \lambda I(\widehat{f} - f^*) \le \rho M^* + \rho(2R) + 2\lambda I(f^*) + \mathcal{E}(f^*).$$

$\square$

**Proof of Theorem 3.** As in the proof of Theorem 1, we begin by defining $t$ as
$$t = \frac{M^*}{M^* + |\widehat{\beta} - \beta^*| + I(\widehat{f} - f^*)},$$

and $\widetilde{f} = t\widehat{f} + (1-t)f^*$ which gives us $|\widetilde{\beta} - \beta^*| + I(\widetilde{f} - f^*) \le M^*$, i.e. $\widetilde{f} \in \mathcal{F}_{local}^0$. Lemma D.1 implies that on the set $\mathcal{T}$ (where $Z_{M^*} \le \rho M^*$) we have
$$\mathcal{E}(\widetilde{f}) + \lambda I(\widetilde{f}) \le Z_{M^*} + \mathcal{E}^* + \lambda I(f^*) \le \rho M^* + \mathcal{E}(f^*) + \lambda I(f^*). \tag{D.5}$$

Be definition of $S_*$ and $I(\cdot)$, we have by the triangle inequality
$$\lambda I(f^*) = \lambda \sum_{j \in S_*} \left\{\|f_j^*\|_n + \lambda P_{st}(f_j^*)\right\} \le \lambda \sum_{j \in S_*} \left\{\|f_j^* - \widetilde{f}_j\|_n + \|\widetilde{f}_j\|_n + \lambda P_{st}(f_j^*)\right\},$$

and by the reverse triangle inequality we have
$$\lambda I(\widetilde{f}) = \lambda \sum_{j \in S_*} \left\{\|\widetilde{f}_j\|_n + \lambda P_{st}(\widetilde{f}_j)\right\} + \lambda \sum_{j \in S_*^c} \left\{\|\widetilde{f}_j\|_n + \lambda P_{st}(\widetilde{f}_j)\right\}$$
$$\ge \lambda \sum_{j \in S_*} \left\{\|\widetilde{f}_j\|_n + \lambda P_{st}(\widetilde{f}_j - f_j^*) - \lambda P_{st}(f_j^*)\right\} + \lambda \sum_{j \in S_*^c} \left\{\|\widetilde{f}_j\|_n + \lambda P_{st}(\widetilde{f}_j)\right\}$$
$$= \lambda \sum_{j \in S_*} \left\{\|\widetilde{f}_j\|_n + \lambda P_{st}(\widetilde{f}_j - f_j^*) - \lambda P_{st}(f_j^*)\right\} + \lambda \sum_{j \in S_*^c} \left\{\|\widetilde{f}_j - f_j^*\|_n + \lambda P_{st}(\widetilde{f}_j - f_j^*)\right\},$$

11

where the last equality follows from the fact that $f_j^* = 0$ for all $j \in S_*^c$. With the above two inequalities combined with (D.5) we get

$$\mathcal{E}(\widetilde{f}) + \lambda \sum_{j \in S_*} \left\{ \|\widetilde{f}_j\|_n + \lambda P_{st}(\widetilde{f}_j - f_j^*) - \lambda P_{st}(f_j^*) \right\} + \lambda \sum_{j \in S_*^c} \left\{ \|\widetilde{f}_j\|_n + \lambda P_{st}(\widetilde{f}_j) \right\}$$
$$\leq \lambda \sum_{j \in S_*} \left\{ \|f_j^* - \widetilde{f}_j\|_n + \|\widetilde{f}_j\|_n + \lambda P_{st}(f_j^*) \right\} + \rho M^* + \mathcal{E}(f^*),$$

which simplifies to

$$\mathcal{E}(\widetilde{f}) + \lambda \sum_{j \in S_*^c} \|\widetilde{f}_j - f_j^*\|_n + \lambda \sum_{j=1}^p \lambda P_{st}(\widetilde{f}_j - f_j^*) \leq \lambda \sum_{j \in S_*} \|f_j^* - \widetilde{f}_j\|_n + 2\lambda^2 \sum_{j \in S_*} P_{st}(f_j^*) + \rho M^* + \mathcal{E}(f^*).$$
(D.6)

Now we add $\lambda |\widetilde{\beta} - \beta^*| + \lambda \sum_{j \in S_*} \|\widetilde{f}_j - f_j^*\|_n$ to both sides of (D.6) to obtain

$$\mathcal{E}(\widetilde{f}) + \lambda \left\{ |\widetilde{\beta} - \beta^*| + I(\widetilde{f} - f^*) \right\} \leq 2\lambda \sum_{j \in S_*} \|f_j^* - \widetilde{f}_j\|_n + \lambda |\widetilde{\beta} - \beta^*| + \rho M^* + \mathcal{E}(f^*) + 2\lambda^2 \sum_{j \in S_*} P_{st}(f_j^*).$$
(D.7)

**Case I.** If

$$2\lambda \sum_{j \in S_*} \|f_j^* - \widetilde{f}_j\|_n + \lambda |\widetilde{\beta} - \beta^*| \leq \rho M^* + \mathcal{E}(f^*) + 2\lambda^2 \sum_{j \in S_*} P_{st}(f_j^*),$$

then (D.7) simplifies to

$$\mathcal{E}(\widetilde{f}) + \lambda \left\{ |\widetilde{\beta} - \beta^*| + I(\widetilde{f} - f^*) \right\} \leq 2\rho M^* + 2\mathcal{E}(f^*) + 4\lambda^2 \sum_{j \in S_*} P_{st}(f_j^*)$$
$$\leq 4\rho M^* \leq 4\frac{\lambda}{8} M^* = \lambda M^*/2,$$

which indicates that $|\widetilde{\beta} - \beta^*| + I(\widetilde{f} - f^*) \leq M^*/2$ which implies that $|\widehat{\beta} - \beta^*| + I(\widehat{f} - f^*) \leq M^*$ and hence we can redo the above arguments and replace $\widetilde{f}$ by $\widehat{f}$.

**Case II.** If instead

$$2\lambda \sum_{j \in S_*} \|f_j^* - \widetilde{f}_j\|_n + \lambda |\widetilde{\beta} - \beta^*| \geq \rho M^* + \mathcal{E}(f^*) + 2\lambda^2 \sum_{j \in S_*} P_{st}(f_j^*),$$

then we have

$$\mathcal{E}(\widetilde{f}) + \lambda \left\{ |\widetilde{\beta} - \beta^*| + I(\widetilde{f} - f^*) \right\} \leq 4\lambda \sum_{j \in S_*} \|f_j^* - \widetilde{f}_j\|_n + 2\lambda |\widetilde{\beta} - \beta^*|. \qquad (D.8)$$

This is equivalent to

$$\mathcal{E}(\widetilde{f}) + \lambda \left\{ \sum_{j \in S_*^c} \|\widetilde{f}_j - f_j^*\|_n + \lambda \sum_{j=1}^p P_{st}(\widetilde{f}_j - f_j^*) \right\} \leq 3\lambda \sum_{j \in S_*} \|f_j^* - \widetilde{f}_j\|_n + \lambda |\widetilde{\beta} - \beta^*|,$$

which means we have that

$$\sum_{j \in S_*^c} \|\widetilde{f}_j - f_j^*\|_n + \lambda \sum_{j=1}^p P_{st}(\widetilde{f}_j - f_j^*) \leq 3 \sum_{j \in S_*} \|f_j^* - \widetilde{f}_j\|_n + |\widetilde{\beta} - \beta^*|,$$

12

and hence by the compatibility condition (D.8) reduces to

$$\mathcal{E}(\widetilde{f}) + \lambda\left\{|\widetilde{\beta} - \beta^*| + I(\widetilde{f} - f^*)\right\} \leq 4\lambda\|\widetilde{f} - f^*\|\sqrt{s_*}/\phi(S_*). \tag{D.9}$$

Since $\widetilde{f}$ and $f^*$ are in $\mathcal{F}^0_{local}$, we invoke the inequality $uv \leq H(v) + G(u)$ to obtain

$$\frac{4\lambda\sqrt{s_*}\|\widetilde{f} - f^*\|}{\phi(S_*)} \leq \frac{8\lambda\sqrt{s_*}}{\phi(S_*)}\left[\frac{\|\widetilde{f} - f^0\|}{2} + \frac{\|f^* - f^0\|}{2}\right]$$

$$\leq H\left(\frac{8\lambda\sqrt{s_*}}{\phi(S_*)}\right) + G\left(\frac{\|\widetilde{f} - f^0\|}{2} + \frac{\|f^* - f^0\|}{2}\right).$$

By the convexity of $G$ and the margin condition we obtain

$$\frac{4\lambda\sqrt{s_*}\|\widetilde{f} - f^*\|}{\phi(S_*)} \leq H\left(\frac{8\lambda\sqrt{s_*}}{\phi(S_*)}\right) + \frac{\mathcal{E}(\widetilde{f})}{2} + \frac{\mathcal{E}(f^*)}{2}.$$

Hence we have

$$\frac{\mathcal{E}(\widetilde{f})}{2} + \lambda\left\{|\widetilde{\beta} - \beta^*| + I(\widetilde{f} - f^*)\right\} \leq H\left(\frac{8\lambda\sqrt{s_*}}{\phi(S_*)}\right) + \frac{\mathcal{E}(f^*)}{2} \leq \rho M^* \leq \lambda M^*/8, \tag{D.10}$$

which implies that $|\widetilde{\beta} - \beta^*| + I(\widetilde{f} - f^*) \leq M^*/2$ which in turn implies $|\widehat{\beta} - \beta^*| + I(\widehat{f} - f^*) \leq M^*$ and hence we can redo the above arguments and replace $\widetilde{f}$ by $\widehat{f}$.

Thus we have shown that

$$\mathcal{E}(\widehat{f}) + \lambda\left\{|\widehat{\beta} - \beta^*| + I(\widehat{f} - f^*)\right\} \leq 4\rho M^*. \tag{D.11}$$

$\square$

# E  The Set $\mathcal{T}$

Theorems 1 and 2 show inequalities holding over the set $\mathcal{T}$. In this section we will show that $\mathcal{T}$ occurs with high probability. This will be shown for the two different types of entropy bounds considered in Section 4. We consider the special case of loss functions linear in $Y_i$ as in Corollaries 1 and 2 and bound the term $\nu_n(f) - \nu_n(f^*)$ in the following theorem.

**Theorem E.1.** *Let $\boldsymbol{x}_i \in \mathbb{R}^p$ and $Y_i \in \mathbb{R}$ denote the fixed covariates and response, respectively, for $i = 1, \ldots, n$. Assume that for any function $f$ the loss $\ell(\cdot)$ is such that*

$$-\ell(f) = -\ell(f, \boldsymbol{x}_i, Y_i) = aY_i f(\boldsymbol{x}_i) + b(f(\boldsymbol{x}_i)), \tag{E.1}$$

*for some $a \in \mathbb{R}\backslash\{0\}$ and function $b : \mathbb{R} \to \mathbb{R}$. Further assume that $Y_i - \mathbb{E}Y_i = Y_i - \mu_i$ are uniformly sub-Gaussian, i.e.*

$$\max_{i=1,\ldots,n} K^2\left(\mathbb{E}e^{(Y_i - \mu_i)^2/K^2} - 1\right) \leq \sigma_0^2, \tag{E.2}$$

*then with probability at-least $1 - 2\exp\left[-n\rho^2 C_1\right] - C\exp\left[-n\rho^2 C_2\right]$ the following inequality holds*

$$\nu_n(f) - \nu_n(f^*) \leq \rho\left[|\beta - \beta^*| + \sum_{j=1}^{p}\|f_j - f_j^*\|_n + \lambda P_{st}(f_j - f_j^*)\right], \tag{E.3}$$

13

for variables $\rho$, $\lambda$ and positive constants $C$, $C_1$, $C_2$ which we specify in the following 3 cases.

**Case 1.** If $\mathcal{F}$ has a logarithmic entropy bound, then the inequality (E.3) holds with $\rho = \kappa \max\left(\sqrt{\frac{T_n}{n}}, \sqrt{\frac{\log p}{n}}\right)$ and $\lambda = 1$ for constants $\kappa = \kappa(a, K, \sigma_0, A_0)$, $C_1 = C_1(K, \sigma_0)$, $C = C(K, \sigma_0)$ and $C_2 = C_2(C, \kappa)$.

**Case 2.** If $\mathcal{F}$ has a polynomial entropy bound with smoothness, then the inequality (E.3) holds with $\rho = \kappa \max\left(n^{-\frac{1}{2+\alpha}}, \sqrt{\frac{\log p}{n}}\right)$ for constants $\kappa = \kappa(a, K, \sigma_0, A_0, \alpha)$, $C_1 = C_1(K, \sigma_0)$, $C = C(K, \sigma_0)$ and $C_2 = C_2(C, \kappa)$. The parameter satisfies $\lambda \asymp \rho$ and $\lambda \geq 8\rho$.

In light of the above theorem, for the case of Theorem 1 where

$$Z_{M^*} = \sup_{I(f-f^*) \leq M^*} |\nu_n(f) - \nu_n(f^*)|,$$

and $\beta, \beta^* \in \mathcal{R}$ where $\mathcal{R}$ is uniformly bounded by $R$ then we have with probability at-least $1 - 2\exp\left[-n\rho^2 C_1\right] - C\exp\left[-n\rho^2 C_2\right]$,

$$Z_{M^*} \leq \rho\left(M^* + 2R\right).$$

In the case of Theorem 3 where

$$Z_{M^*} = \sup_{|\beta - \beta^*| + I(f-f^*) \leq M^*} |\nu_n(f) - \nu_n(f^*)|,$$

we have with probability at-least $1 - 2\exp\left[-n\rho^2 C_1\right] - C\exp\left[-n\rho^2 C_2\right]$,

$$Z_{M^*} \leq \rho M^*.$$

To prove Theorem E.1, we use a few technical lemmas from van de Geer [2000] namely Lemma 8.2 and Corollary 8.3; for the sake of completeness we state these lemmas in Appendix F.

***Proof of Theorem E.1.*** We begin by noting that for any arbitrary function we have

$$\nu_n(f) = (\mathbb{P}_n - \mathbb{P})(-\ell(f)) = \frac{1}{n}\sum_{i=1}^n [aY_i f(\boldsymbol{x}_i) + b(f(\boldsymbol{x}_i))] - \mathbb{E}\left[\frac{1}{n}\sum_{i=1}^n aY_i f(\boldsymbol{x}_i) + b(f(\boldsymbol{x}_i))\right].$$

Since we assume the covariates $\boldsymbol{x}_1, \ldots, \boldsymbol{x}_n$ are fixed we obtain

$$\nu_n(f) = \frac{1}{n}\sum_{i=1}^n a(Y_i - \mu_i)f(\boldsymbol{x}_i) \equiv a\langle \boldsymbol{Y} - \boldsymbol{\mu}, f\rangle_n,$$

where $\mu_i = \mathbb{E}Y_i$. Thus for additive functions $f$ and $f^*$ we obtain

$$\nu_n(f) - \nu_n(f^*) = a\langle \boldsymbol{Y} - \boldsymbol{\mu}, f - f^*\rangle_n = \frac{1}{n}\sum_{i=1}^n a(Y_i - \mu_i)\left[\beta - \beta^* + \sum_{j=1}^p f_j(x_{ij}) - f_j^*(x_{ij})\right]$$

$$= a(\beta - \beta^*)\frac{1}{n}\sum_{i=1}^n (Y_i - \mu_i) + \sum_{j=1}^p \frac{1}{n}\sum_{i=1}^n a(Y_i - \mu_i)(f_j(x_{ij}) - f_j^*(x_{ij}))$$

$$= a(\beta - \beta^*)(\overline{\boldsymbol{Y}} - \overline{\boldsymbol{\mu}}) + \sum_{j=1}^p a\langle \boldsymbol{Y} - \boldsymbol{\mu}, f_j - f_j^*\rangle_n.$$

14

From now on we will assume, without loss of generality, that $|a| = 1$ since this constant is absorbed into a constant $\kappa$ which we define later.

To control the first term, $(\beta - \beta^*)(\overline{Y} - \overline{\mu})$, we simply apply Lemma F.1. For the second part, we consider 2 cases.

**Case 1: Logarithmic Entropy.** We first note that if the entropy bound holds, then the same bound holds (upto a constant) for the class

$$\left\{ \frac{f_j - f_j^*}{\|f_j - f_j^*\|_n + \lambda P_{st}(f_j - f_j^*)} : f_j \in \mathcal{F} \right\}, \tag{E.4}$$

for some $f_j^* \in \mathcal{F}$ for all $j = 1, \ldots, p$. Now we apply Lemma F.2 to the above class by first noting that $R \leq 1$ and then using the bound for Dudley's integral

$$A_0^{1/2} T_n^{1/2} \int_0^1 \log^{1/2}\left(\frac{1}{u} + 1\right) du \leq \widetilde{A}_0 T_n^{1/2}, \tag{E.5}$$

we have for all $\delta$ that satisfy

$$\delta \geq 2C\widetilde{A}_0 \sqrt{\frac{T_n}{n}}, \tag{E.6}$$

where the constant $C$ depends only on $K$ and $\sigma_0$, we have

$$\mathbb{P}\left(\sup_{f_j \in \mathcal{F}} \left|\frac{\frac{1}{n}\sum_{i=1}^n (Y_i - \mu_i)\left(f_j(x_{ij}) - f_j^*(x_{ij})\right)}{\|f_j - f_j^*\|_n + \lambda P_{st}(f_j - f_j^*)}\right| \geq \delta\right) \leq C\exp\left[-\frac{n\delta^2}{4C^2}\right].$$

We can now take $\delta = \rho \geq 2C\widetilde{A}_0 \max\left\{\sqrt{\frac{T_n}{n}}, \sqrt{\frac{\log p}{n}}\right\} \geq 2C\widetilde{A}_0\sqrt{\frac{T_n}{n}}$ which holds for all $\kappa \geq 2C\widetilde{A}_0$. Applying the above result with a union bound gives us

$$\mathbb{P}\left(\max_{j=1,\ldots,p} \sup_{f_j \in \mathcal{F}} \frac{|\langle \boldsymbol{Y} - \boldsymbol{\mu}, f_j - f_j^*\rangle_n|}{\|f_j - f_j^*\|_n + \lambda P_{st}(f_j - f_j^*)} \geq \rho\right)$$
$$\leq pC\exp\left[-\frac{n\rho^2}{4C^2}\right] = C\exp\left[-\frac{n\rho^2}{4C^2} + \log p\right]$$
$$= C\exp\left[-n\rho^2\left(\frac{1}{4C^2} - \frac{\log p}{n\rho^2}\right)\right] \leq C\exp\left[-n\rho^2 C_2\right],$$

for a constant $C_2 > 0$ that depends on $C$ and $\widetilde{A}_0$. To see this, note that

$$\frac{1}{4C^2} - \frac{\log p}{n\rho^2} = \frac{1}{4C^2} - \frac{1}{\kappa \max\left\{\frac{T_n}{\log p}, 1\right\}} \geq \frac{1}{4C^2} - \frac{1}{\kappa},$$

which is positive if $\kappa > 4C^2$. Thus we can take the constant $\kappa$ such that $\kappa > \max\left\{4C^2, 2C\widetilde{A}_0\right\}$. Hence $\kappa$ depends on $C(K, \sigma_0)$ and $\widetilde{A}_0(A_0)$.

**Case 2: Polynomial Entropy with Smoothness.** Now we note that same entropy bound holds for the class

$$\widetilde{\mathcal{F}} = \left\{ \frac{f_j - f_j^*}{\|f_j - f_j^*\|_n + \lambda P_{st}(f_j - f_j^*)} : f_j \in \mathcal{F} \right\}, \tag{E.7}$$

15

and we can now apply Lemma F.2 by noting that
$$\int_0^1 H^{1/2}(u, \widetilde{\mathcal{F}}, Q_n)\, du \leq \widetilde{A}_0 \lambda^{-\alpha/2},$$
for some constant $\widetilde{A}_0 = \widetilde{A}_0(A_0)$. For some $C = C(K, \sigma_0)$ and all $\delta \geq 2C\widetilde{A}_0 \lambda^{-\alpha/2} n^{-1/2}$ we have
$$\mathbb{P}\left(\sup_{f_j \in \mathcal{F}} \frac{|\langle \boldsymbol{Y} - \boldsymbol{\mu}, f_j - f_j^*\rangle_n|}{\|f_j - f_j^*\|_n + \lambda P_{st}(f_j - f_j^*)} \geq \delta\right) \leq C \exp\left[-\frac{n\delta^2}{4C^2}\right]. \tag{E.8}$$
Since $\lambda \geq \rho$ we note that $2C\widetilde{A}_0 \lambda^{-\alpha/2} n^{-1/2} \leq 2C\widetilde{A}_0 \rho^{-\alpha/2} n^{-1/2}$ and that
$$2C\widetilde{A}_0 \rho^{-\alpha/2} n^{-1/2} \leq \rho \Leftrightarrow \rho \geq \left(2C\widetilde{A}_0\right)^{\frac{2}{2+\alpha}} n^{-\frac{1}{2+\alpha}}.$$
Which holds by definition since $\rho = \kappa \max\left(\sqrt{\frac{\log p}{n}}, n^{-\frac{1}{2+\alpha}}\right) \geq \kappa n^{-\frac{1}{2+\alpha}}$ and $\kappa$ is sufficiently large (any $\kappa \geq \left(2C\widetilde{A}_0\right)^{\frac{2}{2+\alpha}}$ would suffice). Therefore, we can take $\delta = \rho$ in (E.8) along with a union bound to obtain
$$\mathbb{P}\left(\max_{j=1,\ldots,p} \sup_{f_j \in \mathcal{F}} \frac{|\langle \boldsymbol{Y} - \boldsymbol{\mu}, f_j - f_j^*\rangle_n|}{\|f_j - f_j^*\|_n + \lambda P_{st}(f_j - f_j^*)} \geq \rho\right) \leq pC \exp\left[-\frac{n\rho^2}{4C^2}\right]$$
$$= C \exp\left[-n\rho^2\left(\frac{1}{4C^2} - \frac{\log p}{n\rho^2}\right)\right]$$
$$\leq C \exp\left[-n\rho^2 C_2\right],$$
for some positive constant $C_2 = C_2(C, \widetilde{A}_0)$ exactly as in Case 1. $\square$

# F  Some Results from van de Geer [2000]

**Lemma F.1** (Lemma 8.2 of van de Geer [2000]). *Suppose that $Y_1 - \mu_1, \ldots, Y_n - \mu_n$ are mean zero sub-Gaussian random variables, i.e., they satisfy (E.2). Then for all $\gamma \in \mathbb{R}^n$ and $\rho > 0$,*
$$\mathbb{P}\left(\left|\sum_{i=1}^n (Y_i - \mu_i)\gamma_i\right| \geq \rho\right) \leq 2\exp\left[-\frac{\rho^2}{8(K^2 + \sigma_0^2)\sum_{i=1}^n \gamma_i^2}\right], \tag{F.1}$$
*in particular if $\gamma_i = 1/n$ then we have*
$$\mathbb{P}\left(\left|\frac{1}{n}\sum_{i=1}^n Y_i - \mu_i\right| \geq \rho\right) \leq 2\exp\left[-\frac{n\rho^2}{8(K^2 + \sigma_0^2)}\right]. \tag{F.2}$$

**Lemma F.2** (Corollary 8.3 of van de Geer [2000]). *Suppose that $\sup_{f_j \in \mathcal{F}} \|f_j\|_n \leq R$ for a univariate function class $\mathcal{F}$ and that $Y_1 - \mu_1, \ldots, Y_n - \mu_n$ are mean zero sub-Gaussian random variables, i.e. they satisfy (E.2). Then for some constant $C = C(K, \sigma_0)$, and for all $\delta > 0$ satisfying*
$$\sqrt{n}\delta \geq 2C\left(\int_0^R H^{1/2}(u, \mathcal{F}, Q_n)\, du \vee R\right), \tag{F.3}$$
*we have*
$$\mathbb{P}\left(\sup_{f_j \in \mathcal{F}}\left|\frac{1}{n}\sum_{i=1}^n (Y_i - \mu_i)f_j(x_{ij})\right| \geq \delta\right) \leq C\exp\left[-\frac{n\delta^2}{4C^2 R^2}\right]. \tag{F.4}$$

16



%

\end{document}